\documentclass[longauth]{aa} 
\usepackage[colorlinks=true, allcolors=blue]{hyperref}

\usepackage{graphicx}
\usepackage{txfonts}
\usepackage{lipsum}
\usepackage{subcaption}        
\usepackage{lscape}            
\usepackage{placeins}
\usepackage{natbib}
\usepackage{graphicx}
\usepackage{subcaption}
\usepackage{wasysym} 
\usepackage{rotating}
\usepackage{multirow} 
\usepackage{xcolor} 
\usepackage{amsmath}
\usepackage{ulem}
\usepackage{blindtext}
\usepackage{color}
\usepackage[title,toc]{appendix}
\usepackage{threeparttable}
\usepackage{amssymb}
\usepackage{longtable}
\usepackage{float}
\usepackage{tabularx}

\usepackage{amsmath}
\usepackage{booktabs}

\newcommand{\msun}{\mbox{M$_{\odot}$}}

\DeclareMathAlphabet{\mathsc}{OT1}{cmr}{m}{sc}
\def\testbx{bx}
\DeclareRobustCommand{\ion}[2]{
\relax\ifmmode
\ifx\testbx\f@series
{\mathbf{#1\,\mathsc{#2}}}\else
{\it{#1\,\mathsc{#2}}}\fi
\else\textup{#1\,{\mdseries\textsc{#2}}}
\fi}
\newcommand{\Ha} {\mbox{H$\alpha$}\,}
\newcommand{\Hb} {\mbox{H$\beta$}\,}

\newcommand{\Nai} {\ion{Na}{i}\,}
\newcommand{\Feii} {\ion{Fe}{ii}\,}
\newcommand{\Caii} {[\ion{Ca}{ii}]\,}

\newcommand{\Cii} {\ion{C}{i}\,}
\newcommand{\Hii} {\ion{H}{ii}\,}
\newcommand{\Hei} {\ion{He}{i}\,}

\newcommand{\Nii} {[\ion{N}{ii}]\,}
\newcommand{\Oi} {[\ion{O}{i}]\,}
\newcommand{\Oii} {[\ion{O}{ii}]\,}

\newcommand{\Sii} {[\ion{S}{ii}]\,}

\newcommand{\Scii} {\ion{Sc}{ii}\,}
\newcommand{\Tiii} {\ion{Ti}{ii}\,}

\newcommand{\Siliconii} {\ion{Si}{ii}\,}

\newcommand{\Caiinforb}{\ion{Ca}{ii}\,}
\newcommand{\Oinforb} {\ion{O}{i}\,}

\usepackage{siunitx}

\newcolumntype{Y}{>{\centering\arraybackslash}X}

\usepackage{graphicx}
\usepackage{txfonts}
\usepackage{lipsum}
\usepackage{subcaption}
\usepackage{lscape}
\usepackage{placeins}
\usepackage{longtable}
\usepackage{graphicx}
\usepackage{CJK}
\usepackage[utf8]{inputenc}
\usepackage{txfonts}
\usepackage{array}
\usepackage{longtable}
\usepackage{hyperref}
\usepackage{tikz}
\usepackage{orcidlink}

\begin{document} 

   \title{SN~2022ngb: A faint, slowly evolving Type IIb supernova with a low-mass envelope}

    \author{
        J.-W. Zhao\inst{\ref{inst1}, \ref{inst1.5}} 
        \and S. Benetti\inst{\ref{inst2}} 
        \and Y.-Z. Cai\inst{\ref{inst3}, \ref{inst4}}\fnmsep\thanks{Corresponding authors: caiyongzhi@ynao.ac.cn (CYZ)}
        \and A. Pastorello\inst{\ref{inst2}}
        \and N. Elias-Rosa\inst{\ref{inst2},\ref{inst5}}
        \and A. Reguitti\inst{\ref{inst2},\ref{inst6}} 
        \and G.~Valerin\inst{\ref{inst2}}
        \and Z.-Y.~Wang\inst{\ref{inst7},\ref{inst8}}
        \and E. Cappellaro\inst{\ref{inst2}}
        \and G.-F. Feng\inst{\ref{inst1}, \ref{inst1.5}}
        \and A. Fiore\inst{\ref{inst9}, \ref{inst2}}
        \and B. Fitzpatrick\inst{\ref{inst10}}
        \and M.~Fraser\inst{\ref{inst10}}
        \and J.~Isern\inst{\ref{inst5}, \ref{inst11}, \ref{inst12}}
        \and E.~Kankare\inst{\ref{inst13}}
        \and T.~Kravtsov \inst{\ref{inst13}, \ref{inst13.5}}
        \and B.~Kumar\inst{\ref{inst1}, \ref{inst1.5}}
        \and P.~Lundqvist\inst{\ref{inst14}}
        \and K.~Matilainen\inst{\ref{inst13}, \ref{inst15}}
        \and S.~Mattila\inst{\ref{inst13}, \ref{inst16}}
        \and P.~A.~Mazzali\inst{\ref{inst16.33}, \ref{inst16.66}}
        \and S.~Moran \inst{\ref{inst17}}
        \and P.~Ochner\inst{\ref{inst2}, \ref{inst20}}
        \and Z.-H. Peng\inst{\ref{inst20.5}}
        \and T.~M.~Reynolds\inst{\ref{inst21}, \ref{inst22}, \ref{inst23}}
        \and I.~Salmaso\inst{\ref{inst24}, \ref{inst2}}
        \and S.~Srivastav\inst{\ref{inst25}}
        \and M.~D.~Stritzinger\inst{\ref{inst26}}
        \and S.~Taubenberger\inst{\ref{inst27}, \ref{inst16.66}}
        \and L.~Tomasella\inst{\ref{inst2}}
        \and J.~Vink\'o\inst{\ref{inst29}, \ref{inst30}, \ref{inst30.33}, \ref{inst30.66}}
        \and J.~C.~Wheeler\inst{\ref{inst30.66}}
        \and S.~Williams\inst{\ref{inst21}, \ref{inst13.5}} 
        \and S.-P. Pei\inst{\ref{inst33}}
        \and Y.-J.~Yang \inst{\ref{inst34}}
        \and X.-K. Liu\inst{\ref{inst1}, \ref{inst1.5}}   
        \and X.-W. Liu\inst{\ref{inst1}, \ref{inst1.5}}\fnmsep\thanks{x.liu@ynu.edu.cn (LXW)}
        \and Y.-P. Yang\inst{\ref{inst1}, \ref{inst1.5}}\fnmsep\thanks{ypyang@ynu.edu.cn (YYP)}
    }
    
    \institute{
        \label{inst1}South-Western Institute for Astronomy Research, Yunnan Key Laboratory of Survey Science, Yunnan University, Kunming, Yunnan 650500, P.R. China
        \and\label{inst1.5}Yunnan Key Laboratory of Survey Science, Yunnan University, Kunming, Yunnan 650500, P.R. China
        \and\label{inst2} INAF - Osservatorio Astronomico di Padova, vicolo dell'Osservatorio 5, I-35122 Padova, Italy
        \and\label{inst3} Yunnan Observatories, Chinese Academy of Sciences (CAS), Kunming 650216, P.R. China
        \and\label{inst4}International Centre of Supernovae, Yunnan Key Laboratory, Kunming 650216, P.R. China
        \and\label{inst5}Institute of Space Sciences (ICE, CSIC), Campus UAB, Carrer de Can Magrans, s/n, E-08193 Barcelona, Spain 
        \and\label{inst6}INAF - Osservatorio Astronomico di Brera, Via E. Bianchi 46, 23807 Merate (LC), Italy 
        \and\label{inst7} School of Astronomy and Space Science, University of Chinese Academy of Sciences, Beijing 100049, P.R. China
        \and\label{inst8} National Astronomical Observatories, Chinese Academy of Sciences, Beijing 100101, P.R. China
        \and\label{inst9} INAF - Osservatorio Astronomico d'Abruzzo, Via Mentore Maggini Snc, 64100 Teramo, Italy
        \and\label{inst10}School of Physics, O'Brien Centre for Science North, University College Dublin, Belfield, Dublin 4, Ireland
        \and\label{inst11}Fabra Observatory, Royal Academy of Sciences and Arts of Barcelona (RACAB), 08001 Barcelona, Spain
        \and\label{inst12}Institute for Space Studies of Catalonia (IEEC), Campus UPC, 08860 Castelldefels (Barcelona), Spain
        \and\label{inst13}Department of Physics and Astronomy, University of Turku, FI-20014 Turku, Finland
        \and\label{inst13.5}Finnish Centre for Astronomy with ESO (FINCA), Quantum, Vesilinnantie 5, University of Turku, FI-20014 Turku, Finland
        \and\label{inst14}The Oskar Klein Centre, Department of Astronomy, Stockholm University, AlbaNova, SE-10691 Stockholm, Sweden
        \and\label{inst15}Nordic Optical Telescope, Aarhus Universitet, Rambla Jos\'e Ana Fern\'andez P\'erez 7, local 5, E-38711 San Antonio, Bre\~na Baja, Santa Cruz de Tenerife, Spain
        \and\label{inst16}School of Sciences, European University Cyprus, Diogenes Street, Engomi, 1516 Nicosia, Cyprus
        \and\label{inst16.33}Astrophysics Research Institute, Liverpool John Moores University, IC2, Liverpool Science Park, 146 Brownlow Hill, Liverpool L3 5RF, UK
        \and\label{inst16.66}Max-Planck-Institut f\"ur Astrophysik, Karl-Schwarzschild Str. 1, D-85741 Garching, Germany
        \and\label{inst17}School of Physics and Astronomy, University of Leicester, University Road, Leicester LE1 7RH, UK
        \and\label{inst20}Universit\'a degli Studi di Padova, Dipartimento di Fisica e Astronomia, Vicolo dell'Osservatorio 2, 35122 Padova, Italy
        \and\label{inst20.5}School of Electronic Science and Engineering, Chongqing University of Posts and Telecommunications, Chongqing 400065, P.R. China
        \and\label{inst21}Tuorla Observatory, Department of Physics and Astronomy, University of Turku, FI-20014 Turku, Finland 
        \and\label{inst22}Cosmic Dawn Center (DAWN)
        \and\label{inst23}Niels Bohr Institute, University of Copenhagen, Jagtvej 128, 2200 K{\o}benhavn N, Denmark
        \and\label{inst24}INAF-Osservatorio Astronomico di Capodimonte, Salita Moiariello 16, 80131 Napoli, Italy
        \and\label{inst25}Astrophysics sub-Department, Department of Physics, University of Oxford, Keble Road, Oxford, OX1 3RH, UK
        \and\label{inst26}Department of Physics and Astronomy, Aarhus University, Ny Munkegade 120, DK-8000 Aarhus C, Denmark
        \and\label{inst27}Technical University of Munich, TUM School of Natural Sciences, Physics Department, James-Franck-Str. 1, 85741 Garching, Germany
        \and\label{inst29}HUN-REN CSFK Konkoly Observatory, MTA Centre of Excellence, Konkoly Thege M. \'ut 15-17, Budapest, 1121, Hungary
        \and\label{inst30}Department of Experimental Physics, Institute of Physics, University of Szeged, D\'om t\'er 9, Szeged, 6720 Hungary
        \and\label{inst30.33}ELTE E\"otv\"os Lor\'and University, Institute of Physics and Astronomy, P\'azm\'any P\'eter s\'et\'any 1A, Budapest 1117, Hungary
        \and\label{inst30.66}Department of Astronomy, University of Texas at Austin, 2515 Speedway, Stop C1400, Austin, TX, 78712-1205, USA
        \and\label{inst33}School of Physics and Electrical Engineering, Liupanshui Normal University, Liupanshui, Guizhou, 553004, P.R. China
        \and\label{inst34} Department of Mathematics and Physics, School of Biomedical Engineering, Southern Medical University, Guangzhou 510515, P.R. China
    }
   \date{Received October 5, 2025; accepted December 9, 2025}
   
  \abstract
   {Type IIb supernovae (SNe IIb) are stellar explosions whose spectra reveal transitional features between hydrogen-rich (Type II) and helium-rich (Type Ib) SNe. Their progenitors are massive stars that were mostly stripped of their hydrogen envelope, likely through binary interaction and/or strong stellar winds. This makes such stars key tools in studies of the late stages of the evolution of massive stars.}
    {We present an extensive photometric and spectroscopic follow-up campaign of the Type~IIb SN~2022ngb. Through the detailed modeling of this dataset, we aim to constrain the key physical parameters of the explosion, infer the nature of the progenitor star and its environment, and probe the dynamical properties of the ejecta.}
    {We analyzed photometric and spectroscopic data of SN~2022ngb. By constructing and modeling the bolometric light curve with semi-analytic models, we were able to estimate the primary explosion parameters. The spectroscopic data were compared with those of well-studied SNe IIb and NLTE models to constrain the properties of the progenitor and the structure of the resulting ejecta.}
    {SN~2022ngb is a low-luminosity SN IIb with a peak bolometric luminosity of $L_{\mathrm{Bol}} = 7.76^{+1.15}_{-1.00} \times 10^{41} \, \mathrm{erg \, s^{-1}}$ and a \textit{V}-band rising time of $24.32\pm0.50$ days. The light curve modeling indicates an ejecta mass of $\sim2.9-3.2$ \msun, an explosion energy of $\sim1.4\times10^{51}\,\mathrm{erg}$, and a low synthesized $^{56}$Ni mass of $\sim0.045$ \msun. The nebular phase spectra exhibit asymmetric line profiles, pointing to a nonspherical explosion and an anisotropic distribution of radioactive material. Our analysis reveals a relatively compact stripped-envelope progenitor with a pre-SN mass of approximately $4.7$ \msun\ (corresponding to a 15$-$16~\msun\ ZAMS star). 
    }
    {Our analysis suggests that SN~2022ngb originated from the explosion of a moderate-mass relatively compact, stripped-envelope star in a binary system. The asymmetries inferred from the nebular phase spectral line features indicate the occurrence of a nonspherical explosion.}

   \titlerunning{SN~2022ngb}
   \maketitle

\nolinenumbers
\section{Introduction} 

Massive stars, with initial masses greater than approximately $8\,\msun$, come to the last phase in their evolution as a spectacular explosion known as a core-collapse supernova (CCSN; \citealt{Heger2003apj, Woosley2002RevModPhys, Janka2012ARNPS}). 
Incapable of generating enough pressure to support itself against its own immense gravity, the core undergoes a catastrophic collapse until it reaches nuclear densities. The subsequent rebound of the core launches a powerful shockwave that travels outwards, disrupting the stellar envelope and releasing a tremendous amount of energy \citep{Woosley1986ARAA}. Observationally, CCSNe are broadly classified based on their spectral features. Supernovae that exhibit prominent hydrogen lines are classified as Type II. Conversely, those that lack hydrogen are collectively referred to as stripped-envelope supernovae (SESNe; \citealt{Clocchiatti1996ApJ}). This group is further subdivided into Type IIb, which exhibit hydrogen lines only at early times; Type Ib, which are characterized by helium lines; and Type Ic, which lack prominent lines of both hydrogen and helium \citep{Filippenko1997ARAA, Modjaz2019NatAs}. The progenitors of SESNe are massive stars that have lost part or all of their outer hydrogen-rich envelope prior to explosion. The envelope stripping is thought to occur primarily through two main channels: powerful stellar winds, such as those from a Wolf-Rayet (WR) star \citep{Smith2014ARAA, Crockett2008MNRAS}, or mass transfer to a companion in a binary system, for instance, through Roche-lobe overflow \citep{Ritter1988AA, Maund2004Natur, Reguitti2025AA}.

Type IIb events belong to a subcategory of SESNe and are considered transitional objects because their early-time spectra are dominated by hydrogen lines. However, in the late phase, the spectra of SNe IIb share similarities with those of Type Ib SNe \citep{Filippenko1988AJ}. Due to the properties of the remaining thin hydrogen envelope, SNe IIb show a variety of light curves and spectra. The subcategories of compact SNe IIb (cIIb) and extended SNe IIb (eIIb) can produce single-peaked and double-peaked light curves, respectively \citep{Chevalier2010ApJL}. In a double-peaked scenario, the first peak is mainly attributed to the cooling phase after shock breakout \citep{Nagy2016aap, Dessart2018AA}, while the secondary peak is powered by radioactive decays \citep{Arnett1982apj,Arnett1989apj} and recombination \citep{Nagy2014AA, Nagy2016aap}. Double-peaked features in the light curves of SNe IIb are common, as exemplified by well-studied objects such as SN~1993J \citep{Richmond1994aj, Barbon1995aaps}, SN~2011fu \citep{Kumar2013mnras}, SN~2013df \citep{Morales-Garoffolo2014mnras,VanDyk2014aj}, SN~2016gkg \citep{Tartaglia2017apjl}, and SN~2024aecx \citep{zou2025arXiv}. In contrast, single-peaked targets (or those with a faint shock cooling phase) seem to be less common, with examples including SN~2008ax \citep{Crockett2008MNRAS, Pastorello2008mnras, Taubenberger2011MNRAS}, SN~2015as \citep{Gangopadhyay2018mnras}, SN~2022crv \citep{Gangopadhyay2023ApJ}, and SN~2024abfo \citep{Reguitti2025AA}. The progenitors of a Type cIIb SNe~are typically characterized by a small radius ($\sim 10^{11} \mathrm{cm}$) and a higher mass before the explosion, whereas the progenitors of Type eIIb SNe~generally have an extended radius ($\sim 10^{13} \mathrm{cm}$) and a lower mass \citep{Barmentloo2024MNRAS}.

In SNe~IIb, both asymmetric explosions and the mixing of ejecta contribute to the diversity in the observed spectra and light curves \citep{Bersten2012ApJ, Jerkstrand2015aap, Fang2024NatAs}. This mixing process is driven by convective \citep{Kifonidis2006A&A} and Rayleigh-Taylor instabilities, the latter of which locally push elements from the interior of the ejecta to the surface \citep{Baal2024MNRAS}. An asymmetric explosion, powered by neutrinos and influenced by intrinsic progenitor characteristics such as spin and magnetic fields \citep{Fang2024NatAs}, leads to a distinct ejecta geometry: a toroidal-like distribution for oxygen and a poloidal distribution for ashes of explosive burning such as calcium. This geometry creates unique, viewing-angle-dependent features in nebular phase spectra, for example, the different emission line shapes for oxygen and calcium.  \cite{Maeda2006ApJ} and \cite{Baal2024MNRAS} found that magnesium and oxygen share a similar distribution, which is consistent with the scenario described above. Calcium, however, originates from explosive burning and thus serves as a more direct probe of the explosion asymmetry.

In this paper, we present an analysis of the photometric and spectroscopic observations and theoretical modeling of SN~2022ngb, a Type IIb SN in UGC 11380. Section \ref{sec_basicinfo} provides the basic parameters of SN~2022ngb and outlines the data reduction procedures. In Sect. \ref{sec_photometric}, we analyze the photometric evolution and apply a simple model based on a modified Arnett formalism and a two-component structure \citep{Arnett1982apj,Arnett1989apj,Nagy2014AA,Nagy2016aap}. The spectroscopic evolution is detailed in Sect. \ref{sec_spec}, where we compare the spectra of SN~2022ngb with those of other SNe IIb. Furthermore, we describe our use of  synthetic spectra generated by NLTE models \citep{Jerkstrand2015aap,Barmentloo2024MNRAS,Dessart2016MNRAS} to constrain the progenitor. A detailed discussion of the physical properties of the progenitor, the explosion, and the resulting remnant is presented in Sect. \ref{sec_phys}. Finally, we summarize our findings in Sect. \ref{sec_con}.

\section{Distance, reddening, and host galaxy}
\label{sec_basicinfo}

\begin{figure}[htb]
\centering
\includegraphics[width=9cm]{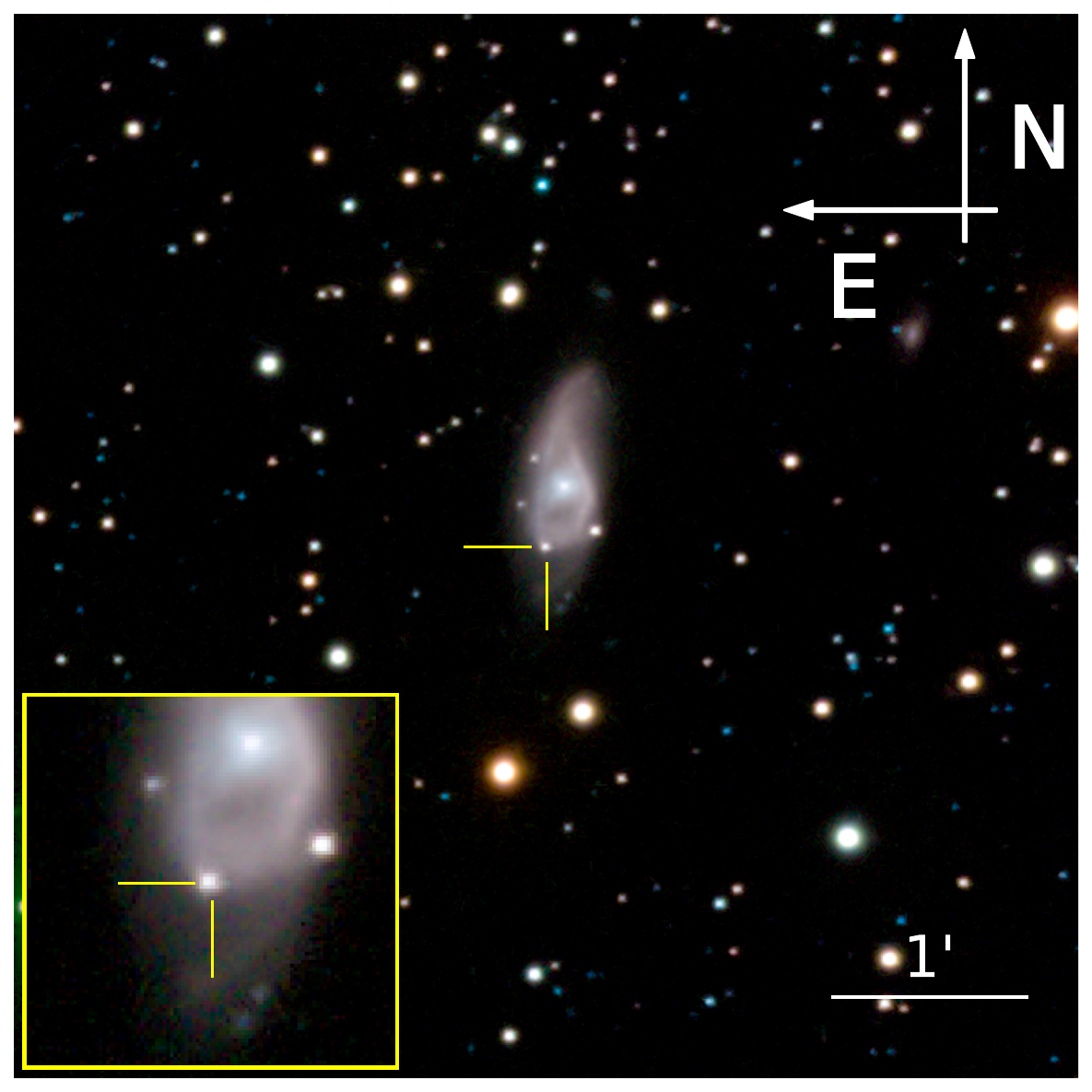}
\caption{Composite \textit{BVr}-band image constructed from images obtained with the NOT/ALFOSC. The location of SN~2022ngb with the host galaxy UGC~11380. 
}
\label{phot_image}
\end{figure}

SN~2022ngb (ATLAS22res) was first reported by the Asteroid Terrestrial-impact Last Alert System \citep[ATLAS;][]{Tonry2018apj, Tonry2018pasp, Smith2020pasp} on June 21, 2022, corresponding to MJD 59751.48. The discovery magnitude was 18.878 in the ATLAS orange (\textit{o}) band. The J2000 coordinates of the transient are $\mathrm{RA} = 18^{h}56^{m}51^{s}.48$ and $\mathrm{DEC}=+36^{\circ}37'07''.82$ \citep{Tonry2022TNSTR1729}. 
SN~2022ngb exploded in UGC 11380, 18\farcs29 south and 5\farcs27 east from the galaxy center (see Fig.~\ref{phot_image}).

Making comparisons between an early spectrum of SN~2022ngb and archival spectra using the SNID tool \citep{Blondin2007ApJ} a Type IIb SN classification was initially proposed by \citet{Izzo2022TNSAN}. 
According to SIMBAD \citep{Wenger2000AAS} and the NASA/IPAC Extragalactic Database (NED; \citealt{Helou1991ASSL}), UGC 11380 is an Sab galaxy \citep{deVaucouleurs1991book}. It has a redshift of $z = 0.00965 \pm 0.00006$ \citep{Masters2014MNRAS}. The distance to the galaxy, derived using the Tully-Fisher relation \citep{Tully2016aj}, is $32.2 \pm 2.8$ Mpc \citep{Kourkchi2020ApJ}, which corresponds to a distance modulus of $\mu = 32.5 \pm 0.2$ mag.

The total line-of-sight reddening, $E(B-V)_{\mathrm{Total}}$, is the sum of the foreground component from the Milky Way ($E(B-V)_{\mathrm{MW}}$) and the internal component from the host galaxy ($E(B-V)_{\mathrm{host}}$). The Milky Way (MW) foreground reddening is $E(B-V)_{\mathrm{MW}} = 0.085\, \mathrm{mag}$ \citep{Schlafly2011apj}. To estimate the host galaxy reddening, we measured the equivalent width (EW) of the Na \textsc{i} D absorption line in our early spectra ($-$21.1 days from \textit{V}-band maximum). The measured Na \textsc{i} D EW at the host galaxy redshift is found to be very similar to that of the MW foreground absorption (both are around $\sim3.3$ \AA), thus $E(B-V)_{\mathrm{host}} \approx 0.085\, \mathrm{mag}$. This yields an approximate total reddening of $E(B-V)_{\mathrm{Total}} \approx 0.170\, \mathrm{mag}$. A summary of these properties is presented in Table \ref{tab_targetandhostinfo}.

\begin{table}[htb]
\centering
\caption{Information of SN 2022ngb and the host galaxy UGC~11380.}
\label{tab_targetandhostinfo}
\begin{tabular}{llc}
\hline
\hline
Item                         &   Value                             & Source           \\ 
\hline
\hline
SN 2022ngb:                      &                                     &                     \\    
\quad RA (J2000)                   &   $18^{h}56^{m}51^{s}.48$          & 1                   \\   
\quad DEC (J2000)                  &   $+36^{\circ}37'07''.82$          & 1                   \\ 
\quad Explosion epoch (MJD)       &   $59749.9\pm0.5$                          & Section \ref{sec_photometric} \\
\quad Discovery date (MJD)      &   59751.48                          & 1                   \\
\quad Type                        &   IIb                               & 2                   \\
\quad $E(B-V)_{\rm{MW}}$               &   0.085 mag                         & 3                   \\
\quad $E(B-V)_{\rm{host}}$             &   0.085 mag                         & Section \ref{sec_basicinfo}                \\ 
\\

UGC 11380:                       &                                     &                     \\
\quad Type                        &   Sab                               & 4                   \\
\quad Redshift                    &   $0.00965 \pm 0.00006$             & 5                   \\
\quad Distance                    &   $32.2 \pm 2.8$ Mpc              & 6                   \\
\quad Distance modulus            &   $32.5 \pm 0.2$ mag       & 6                   \\ 
\hline
\hline
\end{tabular}
\vspace{1em}
\scriptsize 
\raggedright

\textbf{NOTE:} 1 = \citet{Tonry2022TNSTR1729}; 2 = \citet{Izzo2022TNSAN}; 3 = \citet{Schlafly2011apj}; 4 = \url{https://ned.ipac.caltech.edu/} ; 5 = \citet{Masters2014MNRAS}; 6 = \citet{Kourkchi2020ApJ}.
\end{table}

\begin{figure*}[htbp]
\centering
\includegraphics[width=1.0\linewidth]{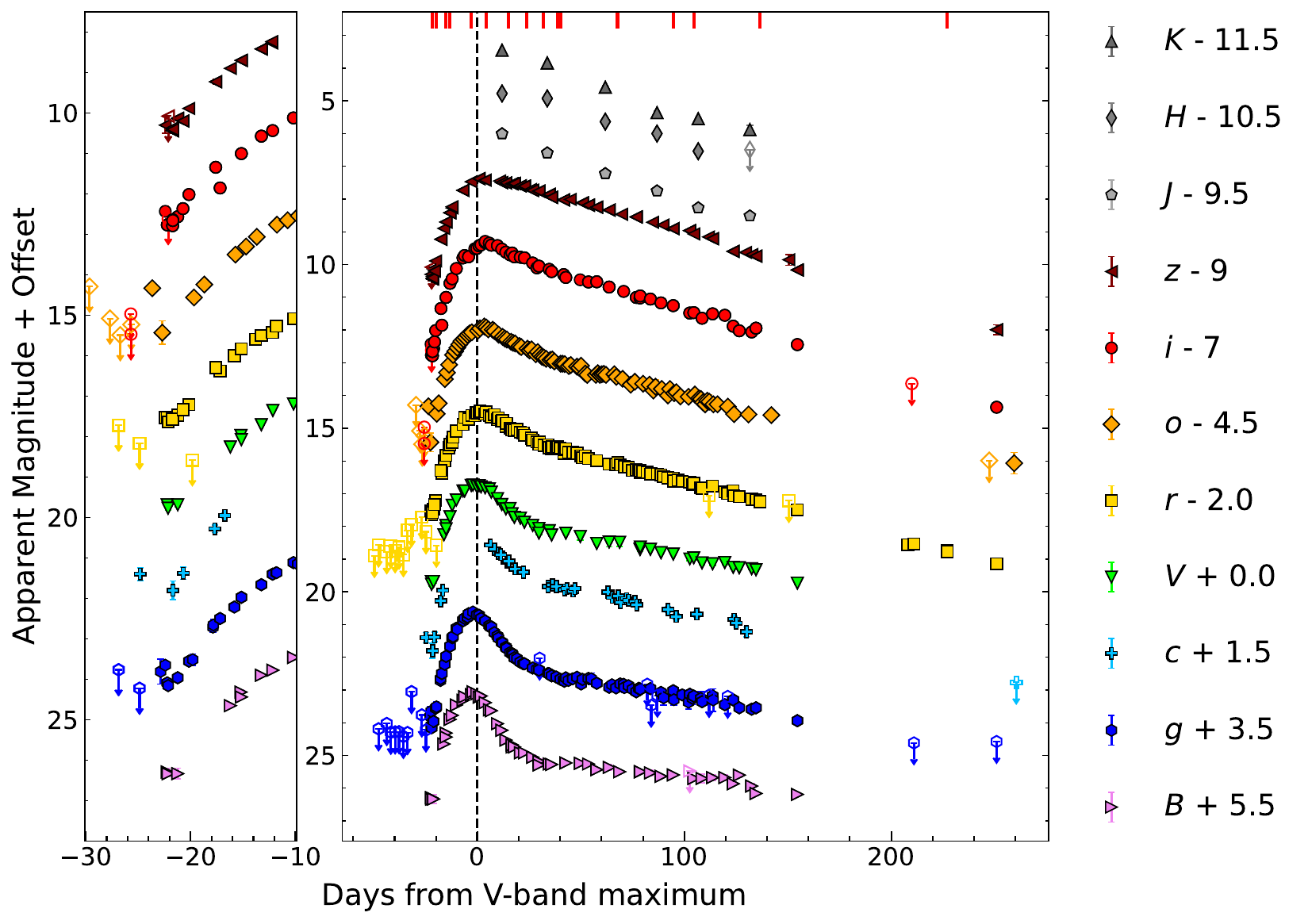}
\caption{Multiband (optical \textit{BgcVroiz} and NIR \textit{JHK}) light curves of SN~2022ngb, showing apparent magnitudes with arbitrary offsets. Different colors and symbols correspond to different photometric filters, as indicated in the legend. The left panel displays the full light curve evolution, while the right panel provides a zoomed-in view of the early-phase light curves. All photometric points include error bars that, in general, are smaller than the marker sizes. }
\label{lcurve}
\end{figure*}

\section{Photometric analysis}
\label{sec_photometric}
\subsection{Apparent magnitude light curves}
\label{sub_applc}

We obtained optical and near-infrared (NIR) photometric data for SN~2022ngb. The observations were carried out in the optical \textit{BgcVroiz} bands and the NIR \textit{JHK} bands. The photometric data  reduction process is described in Appendix~\ref{app_datareduction}. The resulting apparent magnitude light curves are shown in Fig.~\ref{lcurve}, as for \textit{BVJHK} bands, we took advantage of Vega mags, whereas we used AB mags for the \textit{gcvroiz} bands. Our optical monitoring of SN~2022ngb spanned approximately 280 days, yielding well-sampled light curves. In contrast, the NIR observations covered only the declining phase. Using observations from the ATLAS survey, we constrained the explosion epoch to be the midpoint between the last nondetection (an upper limit in the \textit{o}-band) at MJD 59749.52 and the first detection at MJD 59750.38. This yields an explosion epoch of MJD $59749.9\pm0.5$. A summary of the features of the apparent light curve is provided in Table \ref{tab_appapplc}.

The optical light curves enable us to obtain a precise determination of the peak parameters in each band. To this end, we fit the light curve around the time of maximum brightness in each band using a Legendre polynomial. In the \textit{V} band, the epoch of maximum light is determined to be MJD $59774.22 \pm 0.06$, with a peak apparent magnitude of $m_{V}=16.74 \pm 0.01$. Accounting for the explosion epoch (MJD $59749.9\pm0.5$), a rise time (from explosion epoch to radioactive powered peak) in the $V$ band of $24.32 \pm 0.50$ days was estimated. This value is slightly longer than the rise times of other SNe IIb 
such as SN~2008ax (20.7 days; \citealp{Pastorello2008mnras}), SN~2011fu (23.4 days; \citealp{Morales-Garoffolo2015mnras}), SN~2013df (20.15 days; \citealp{Morales-Garoffolo2014mnras}), and SN~2024abfo (22.1 days; \citealp{Reguitti2025AA}). 
The same fitting procedure was applied to all other optical bands, revealing a trend in which bluer bands have longer rise times (see Table \ref{tab_appapplc}).

The early-time light curve of SN~2022ngb exhibits a rapid initial decline in the ATLAS \textit{c} and \textit{o} bands. This feature is identified as the shock breakout cooling emission, frequently observed in SNe IIb.
As shown in the right panel of Fig.~\ref{lcurve}, this decline phase lasts approximately 3 days. While this phenomenon has been observed in other SNe IIb such as SN~1993J \citep{Richmond1996aj}, SN~2011fu \citep{Kumar2013mnras}, and SN~2013df \citep{Szalai2016mnras}, the shock cooling emission of SN~2022ngb is notably faint, similarly to SN~2022crv \citep{Gangopadhyay2023ApJ} and SN~2024abfo \citep{Reguitti2025AA}. This implies that the progenitor of SN~2022ngb possessed likely a compact and thin envelope. This case is also distinct from SNe that show marginally detectable shock cooling peak such as SN~2008ax \citep{Pastorello2008mnras, Roming2009ApJ} and SN~2020acat \citep{Medler2022MNRAS, Ergon2024A&A}, which are thought to originate from progenitors that were almost completely stripped of their hydrogen envelopes.

We calculated the post-maximum decline rates of the apparent light curves in each optical band using a simple linear fit. The rate of decline was measured over two distinct epochs: the first 15 days ($\gamma_{0-15}$) and the period between 15 and 100 days ($\gamma_{15-100}$). The resulting rates, listed in Table \ref{tab_appapplc}, show a clear wavelength dependence: bluer bands such as \textit{B} and \textit{g} decline more steeply than redder bands, such as \textit{V}, \textit{r}, \textit{o}, \textit{i}, and \textit{z}. This trend is a known consequence of the temperature evolution of the supernova ejecta. As the photosphere cools, its blackbody-like emission peak shifts towards redder wavelengths, leading to a more pronounced drop in flux in the blue part of the spectrum. This physical process is entirely consistent with the observed reddening in the color evolution.

At a later phase (from 15 to 100 days), the trend  is reversed. The decline in the blue bands becomes significantly slower and the decline rate systematically increases with wavelength, from 0.77 $\mathrm{mag \times (100\,d)^{-1}}$ in the \textit{B} band to 1.94 $\mathrm{mag \times (100\,d)^{-1}}$ in the \textit{z} band. \footnote{The NIR bands were excluded from the analysis of the decline rates because of their poor sampling.} This transition marks the period in which the photosphere rapidly recedes through the ejecta in the co-moving frame, driven by widespread recombination \citep{Nagy2014AA}. The ejecta become increasingly transparent, especially at shorter wavelengths. This allows photons powered by the radioactive decay of $^{56}$Co in the deeper, inner regions to escape, thus slowing the photometric decline in the blue bands \citep{Kumar2013mnras}. Meanwhile, the faster decline in the red bands also indicates the potential opening of efficient cooling channels through growing nebular emission lines in the late transition phase \citep{Baal2023MNRAS}. 

\subsection{Absolute light curves}
\begin{figure}[htb]
\centering
\includegraphics[width=1.0\linewidth]{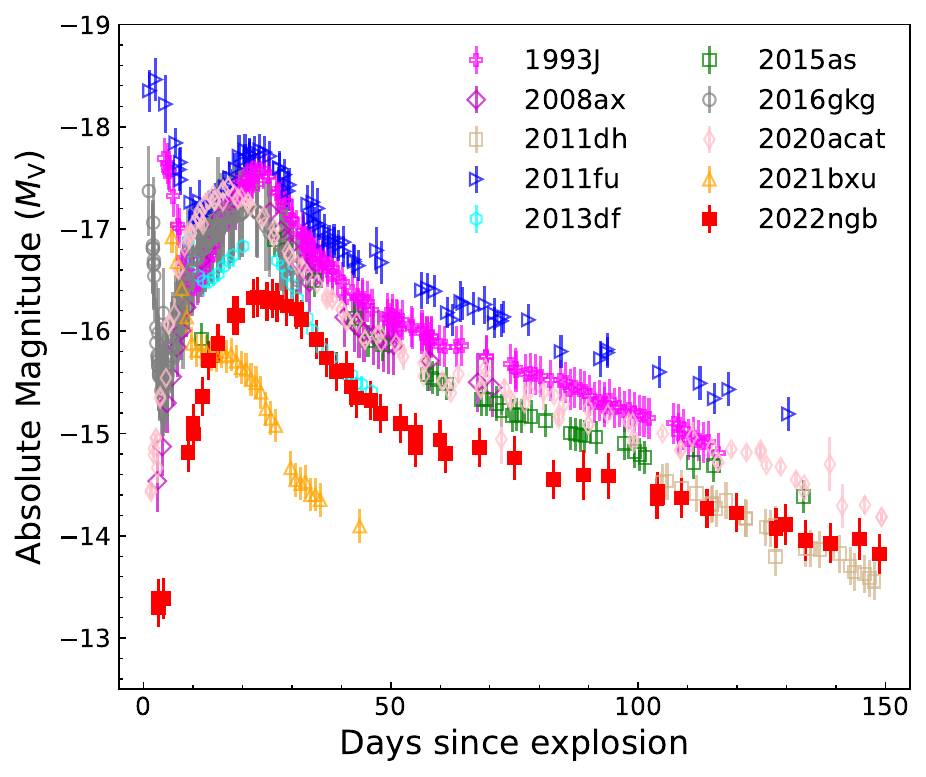}
\caption{Absolute \textit{V}-band light curve of SN~2022ngb compared with other SNe IIb.  
All light curves have been corrected for reddening and shifted according to the distances listed in Table \ref{tab_appSNIIbInfo}.}
\label{phot_absolutemagnitudecmp}
\end{figure}

SNe IIb exhibit generally consistent luminosities of the radioactive decay powered peak, in the range between $-$16.5 mag and $-$18 mag \citep{Taddia2018A&A, Stritzinger2018A&A}.
This diversity is attributed to the wide range of ejecta parameters. We applied the reddening correction and distance modulus to the apparent light curves of SN~2022ngb to derive the absolute light curves. The peak absolute magnitude of SN~2022ngb in the \textit{V} band is $M_V = -16.33 \pm 0.19$ mag. Figure~\ref{phot_absolutemagnitudecmp} presents the absolute \textit{V}-band light curves of SN~2022ngb and a sample of comparison SNe. It is evident that the peak absolute magnitude of SN~2022ngb is fainter than of other SNe IIb, such as SN~1993J ($-17.57 \pm 0.24$ mag; \citealp{Richmond1994aj}), SN~2008ax ($-17.61 \pm 0.43$ mag; \citealp{Morales-Garoffolo2014mnras}), SN~2011dh ($-17.12 \pm 0.18$ mag; \citealp{Sahu2013mnras}), SN~2011fu ($-18.50 \pm 0.24$ mag; \citealp{Kumar2013mnras}), and SN~2022acat ($-17.62 \pm 0.01$ mag; \citealp{Medler2022MNRAS}). In contrast, the absolute light curve of SN~2022ngb resembles those of less luminous objects, such as SN~2013df ($-16.85 \pm 0.08$ mag; \citealp{Szalai2016mnras}), SN~1996cb ($-16.22$ mag; \citealp{Qiu1999aj}), SN~2011ei ($-16.0$ mag; \citealp{Milisavljevic2013apj}), and SN~2024abfo ($-16.32$ mag; \citealp{Reguitti2025AA}). This suggests that the radioactive energy input of SN~2022ngb is lower than that of the more luminous events. 

\subsection{Color evolution}
\begin{figure*}[htbp]
\centering
\includegraphics[width=1.0\linewidth]{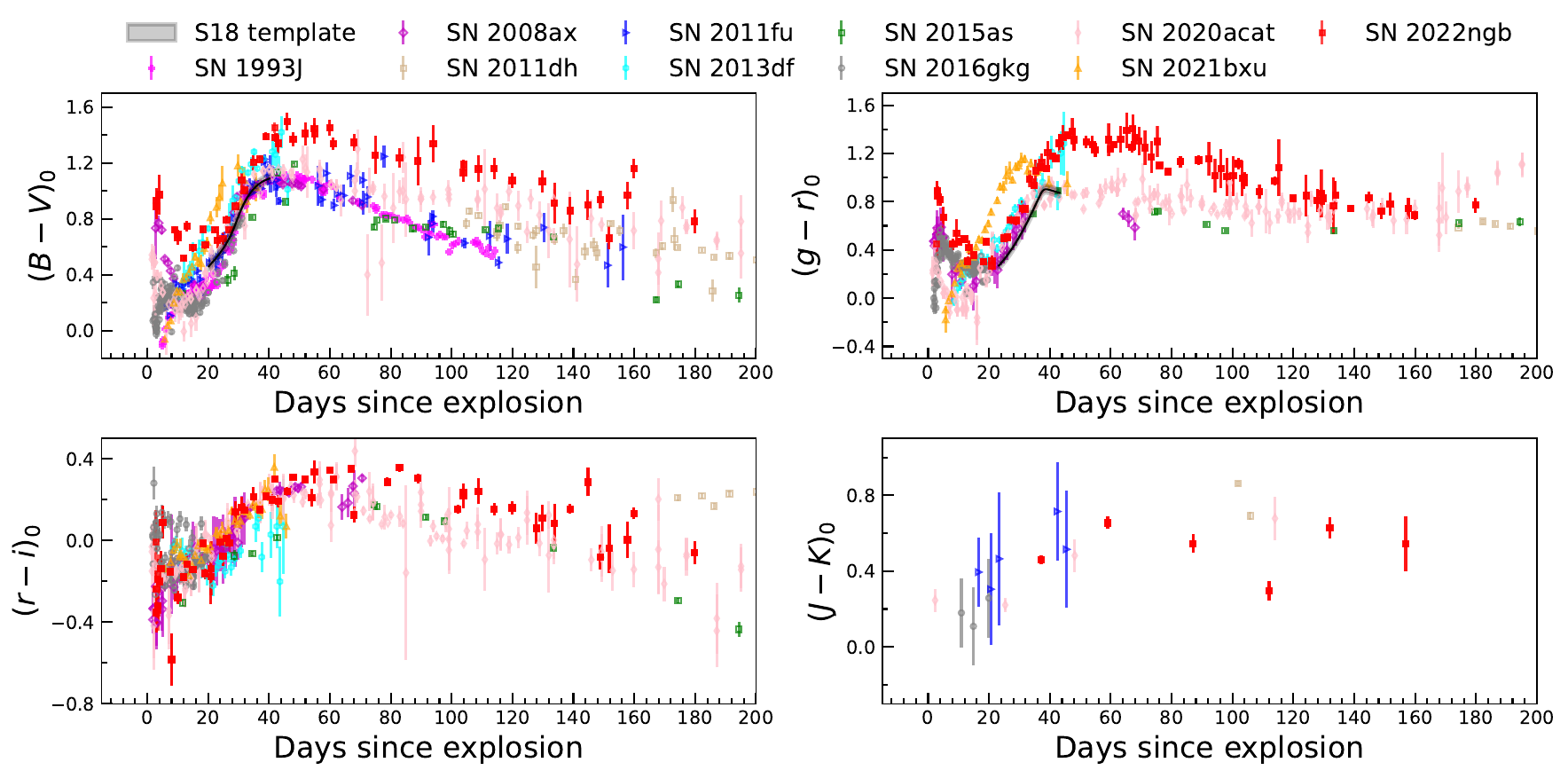}
\caption{Intrinsic color evolution of SN~2022ngb, compared with a sample of SNe~IIb. The color curves are corrected for a total line-of-sight extinction. The black lines over-plotted with gray area around represent the color evolution template from \cite{Stritzinger2018A&A}.}
\label{color_curve}
\end{figure*}

The intrinsic color evolution of SN~2022ngb, presented in Fig.~\ref{color_curve}, is compared with those of SN~1993J, SN~2008ax, SN~2011dh, SN~2011fu, SN~2013df, SN~2015as, SN~2016gkg, SN~2020acat, SN~2021bxu, and the color evolution template of SNe IIb from \citet[hereafter S18]{Stritzinger2018A&A}. All color curves have been corrected for reddening using the parameters listed in Table \ref{tab_targetandhostinfo}. 

The intrinsic $(B-V)_0$ and $(g-r)_0$ colors of SN~2022ngb were initially redder than those of the comparison sample. During the first two weeks after the explosion, the color of SN~2022ngb evolved rapidly to bluer colors: the $(B - V)_0$ index decreased from $0.92\pm0.10$ mag to $0.52\pm0.02$ mag, $(g - r)_0$ from $0.90\pm0.08$ mag to $0.45\pm0.05$ mag, and $(r - i)_0$ from  $0.09\pm0.07$ mag to $-0.58\pm0.12$ mag. Following this initial blue-ward trend, the color evolution reversed, becoming progressively redder and reaching a peak at approximately 50 days past explosion, which corresponds to the start of the plateau in light curve. At this epoch, the $(B - V)_0$ and $(g - r)_0$ indices reached approximately 1.5 mag, while $(r - i)_0$ peaked at about 0.3 mag and later declined again, indicating a renewed evolution toward bluer colors. By approximately 160 days post-explosion, the indices had fallen to $(B-V)_0=0.66\pm0.08$ mag, $(g-r)_0=0.77\pm0.06$ mag, and $(r-i)_0= -0.06\pm0.05$ mag. We note that the decline rate of $(r-i)_0$, approximately $0.003\,\mathrm{mag\,day^{-1}}$, is slower than those of $(B-V)_0$ and $(g-r)_0$ ($\sim0.005\,\mathrm{mag\,day^{-1}}$ for both). The $(B-V)_0$ and $(g-r)_0$ colors of SN~2022ngb is slightly redder than the template of \cite{Stritzinger2018A&A}, but generally consistent with the trend of the template, especially in the early time.

The initial evolution to bluer colors followed by a red-ward trend is characteristic of a brief shock-cooling phase succeeded by heating from the decay of radioactive material. This behavior suggests that SN~2022ngb originated from a relatively compact progenitor, similar to other SNe IIb such as SN~2008ax, SN~2020acat, SN~2022crv, and SN~2024abfo. The post-maximum color evolution indicates that the transition from the photospheric to the nebular phase began at around 40$-$60 days past explosion, as the helium-rich, radioactively heated core gradually became visible. 

Despite the sparse NIR photometric sampling, the color evolution is also tentatively inferred in this domain. 
The $(J-K)_0$ color initially follows a similar trend as the optical colors. However, at later phases, it reddens again starting from $\sim140$ days. This late-time color evolution can be likely explained by an IR echo from pre-existing dust surrounding the SN. An alternative explanation is the potential formation of dust, which could also contribute to the NIR bands.

\subsection{Pseudo-bolometric light curves}
To construct the pseudo-bolometric light curve, we first correct the apparent magnitude in each band for reddening and then converted the corrected magnitudes to fluxes with the corresponding distance. The parameters for this conversion are taken from Table \ref{tab_targetandhostinfo} and the reddening correction is applied using the method described in the previous section. We then used the \texttt{SuperBol} code, as detailed in \citet{Nicholl2018RNAAS}, to construct the pseudo-bolometric light curve.

\begin{figure}[htbp]
\centering
\includegraphics[width=1.0\linewidth]{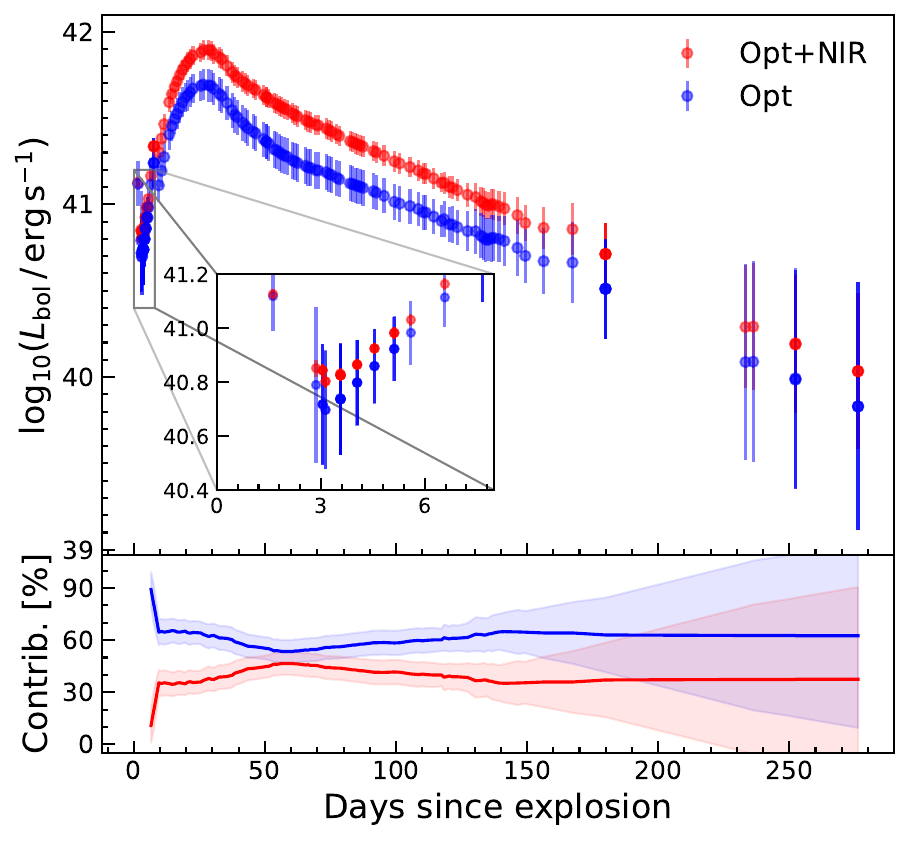}
\caption{Pseudo-bolometric light curve of SN~2022ngb. Top panels: Light curve constructed using optical and NIR data only. Bottom panels: Evolution of the contribution of the individual wavelength ranges (blue corresponds to the optical and the red to NIR) with time of the pseudo-bolometric luminosity. The blue points stand for the pseudo-bolometric light curve solely constructed using optical data, while the red points represent the pseudo-bolometric light curve constructed with optical and NIR data.
}
\label{bol_lc}
\end{figure}

The pseudo-bolometric flux at each epoch is calculated by integrating the monochromatic fluxes over the optical (\textit{BgcVroiz}) and NIR (\textit{JHK}) bands. The resulting pseudo-bolometric light curve is presented in Fig.~\ref{bol_lc}. The uncertainty of the pseudo-bolometric flux was propagated accounting for the error of each photometric point and the error introduced by the interpolation. The complete pseudo-bolometric light curve is constructed by combining two separate segments: the first covers the period from MJD 59751 to 59925 and the second covers from MJD 59925 onward. To ensure that all bands have flux measurements at common epochs, we perform a linear interpolation for each band with respect to a reference band. Due to data availability, different reference bands are used for the two segments. For the first segment, corresponding to the shock cooling phase, the \textit{o} band serves as the reference due to the lack of \textit{r}-band data. For the later segment, the \textit{r} band is used as it provides better temporal coverage than the \textit{o} band.

A significant limitation in the construction of our pseudo-bolometric light curve is the lack of UV data. As noted by \cite{Arcavi2022apj}, the absence of UV photometry can lead to substantially underestimate the total bolometric flux, particularly during the early phases. This issue is highlighted in recent studies of CCSNe. For example, research on SN~2024ggi \citep{Chen2024apjl} and SN~2023ixf \citep{Teja2023apjl} suggests that both UV and NIR bands can contribute significantly to the total flux. The exclusion of these bands can also lead to a lower temperature in the proximity of the maximum light. Consequently, our calculated pseudo-bolometric luminosity of SN~2022ngb at maximum, $L_{\mathrm{peak}} \sim 7.85 \times 10^{41}~\mathrm{erg~s^{-1}}$, should be  regarded as a lower limit to the real luminosity. Furthermore, the impact of the missing UV contribution is time-dependent. The study on SN~2020acat by \citet{Medler2022MNRAS} showed an  UV contribution of about 30\% during the early stages. However, the same study indicated that this contribution becomes negligible after approximately 50 days. This finding implies that our luminosity estimates for SN~2022ngb can be largely underestimated during the early phases of its evolution.

The sparse temporal coverage of our NIR data, particularly during the early phases and around the peak brightness, introduces an additional source of uncertainty in the construction of our pseudo-bolometric light curve. For the early stages, studies of SN~2020acat indicate that the NIR bands contribute little to the total pseudo-bolometric flux. However, extrapolating from later epochs to fill these early gaps may lead to an overestimation of the pseudo-bolometric flux. Conversely, in the later stages of the evolution, the NIR contribution increases dramatically, reaching approximately 40\%. This means that at late times, the bolometric flux becomes highly dependent on the NIR measurements. In this case, extrapolation of the sparse late-time NIR data can introduce significant errors in the pseudo-bolometric flux. As a result, the bolometric data at late stages (phase > 140 days) should be used with caution when deriving physical properties, as they may lead to biased results.

An additional pseudo-bolometric light curve, constructed using only optical data, is also provided in Fig.~\ref{bol_lc}. The individual contributions of the optical and NIR bands to the full pseudo-bolometric light curve are shown in the lower panel. For the earliest stages, the NIR contribution could not be calculated due to a lack of data. As shown in the figure, the optical bands dominate the bolometric flux throughout the evolution of SN~2022ngb, particularly in the early stages. The NIR contribution increases from approximately 10\% to a peak of nearly 45\% at around 60 days, and then begins to decrease slowly. This evolutionary trend is similar to that observed in SN~2020acat \citep{Medler2022MNRAS}.

\begin{figure}[htbp]
\centering
\includegraphics[width=1.0\linewidth]{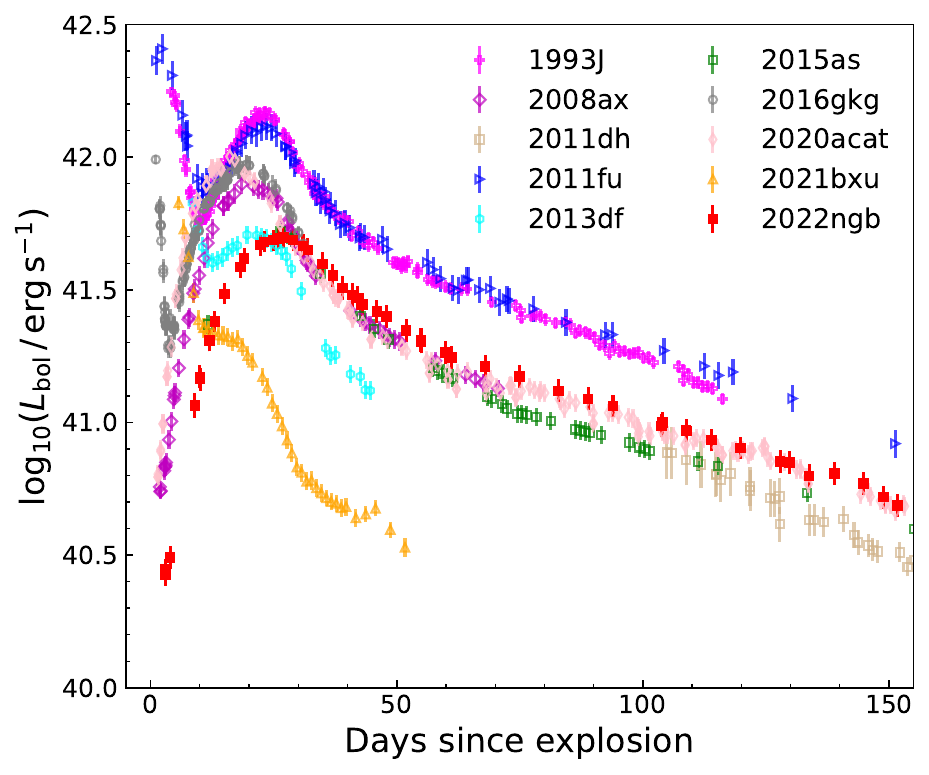}
\caption{Pseudo-bolometric light curve of SN~2022ngb, shown alongside a sample of SNe IIb. To ensure the validity of the comparison, the pseudo-bolometric light curves for all events were constructed by integrating the flux only over the photometric bands common to the entire sample.}
\label{bol_lc_compare}
\end{figure}

Figure \ref{bol_lc_compare} shows the pseudo-bolometric light curves of SNe 2008ax, 2011dh, 2011fu, 2013df, 2015as, 2016gkg, 2020acat, and 2021bxu for comparison with SN~2022ngb. The pseudo-bolometric luminosities for each SN were estimated using only optical bands (from \textit{B} to \textit{z}) to ensure a consistent comparison by excluding the influence of the NIR and UV bands. This comparison reveals a wide range of evolutionary behaviors within the sample, in particular in the peak luminosity. SN~2022ngb exhibits a relatively low peak luminosity, indicating a smaller $M_{\text{Ni}}$ value. Furthermore, its evolution closely follows those of SN~2020acat, SN~2015as, and SN~2008ax, all of which display light curves with a faint or absent early shock-cooling phase. In contrast, SNe IIb such as SN~2011fu, SN~2013df, and SN~2016gkg show prominent double-peaked features, suggesting  a more evident contribution from the expanded progenitor's envelope for these objects than for SN~2022ngb.

\subsection{Light curve modeling} \label{LCM}
The bolometric light curve of SN~2022ngb, constructed by applying a black-body (BB) correction to the pseudo-bolometric light curve. Using the data of each bands to fit the SED of black-body, it was modeled using both an Arnett-like approximation \citep{Arnett1982apj, Arnett1989apj, Chatzopoulos2012apj} and the two-component \texttt{LC2} model \citep{Nagy2016aap}. Following the methodology applied to SN~2015as \citep{Gangopadhyay2018mnras} and SN~2024abfo \citep{Reguitti2025AA}, we assumed the SN~is powered by the $^{56}\text{Ni}$$\rightarrow$ $^{56}\text{Co}$$\rightarrow$ $^{56}\text{Fe}$ radioactive decay chain during the main luminosity peak. Under this assumption, the bolometric light curve can be described by Arnett's model. As suggested by \cite{Gangopadhyay2018mnras}, the mass of $^{56}\text{Ni}$ can be estimated from the peak luminosity \citep{Prentice2016mnras} and the rise time of the bolometric light curve can serve as an approximation for the diffusion timescale. Based on the observational data for SN~2022ngb, we first applied Arnett's rule to estimate the initial model parameters. For this calculation, we assumed an idealized scenario: the ejecta expansion is spherically symmetric and homologous, and the opacity is constant \citep{Arnett1982apj}. We also assume that the radioactive material is located at the center of the ejecta, without any hydrodynamic mixing \citep{Arnett1989apj, Arnett1982apj}. Under these simplifying assumptions, the diffusion time $\tau_{m}$ and kinetic energy can be expressed as
\begin{equation}
\tau_{\rm{m}} = \sqrt{\frac{2 \kappa M_{\rm{ej}}}{\beta c v_{\rm{ph}}}}, \;\;\; E_{\rm{k}} = 0.3 M_{\rm{ej}} v_{\rm{ph}}^2 \, .
\end{equation}
Here, $\beta = 13.8$ is the integration constant in Arnett's model \citep{Arnett1982apj, Valenti2008ApJ}. Following the methodology applied to SN~2015as, we assumed the diffusion timescale ($\tau_{\rm{m}}$) is coincident with the observed bolometric rise time; hence, $\tau_{\rm{m}} = 28.96 \, \mathrm{days}$. The opacity, $\kappa$, is assumed to be dominated by electron scattering. For the He-rich ejecta of SNe IIb, calibrations against \texttt{SNEC} hydrodynamic models suggest an average opacity of $0.19 \pm 0.01 \, \mathrm{cm^2 \, g^{-1}}$ \citep{Nagy2018ApJ, Nagy2016aap}. We therefore adopt $\kappa$ values in the range between $0.18$ and $0.2 \, \mathrm{cm^2 \, g^{-1}}$. These inputs yield a derived ejecta mass in the range from $M_\mathrm{ej} = 2.91 \, \mathrm{M_{\odot}}$ to $3.23 \, \mathrm{M_{\odot}}$. The corresponding kinetic energy is $E_k \approx 1.4 \times10^{51}\,\mathrm{erg}$. We noted that the longer rise time of SN~2022ngb suggests a larger ejecta mass, as this would increase the photon diffusion timescale. For comparison, previous studies of SNe (e.g., based on SN~2015as and SN~2020acat) adopted a lower opacity range from $0.06 \, \mathrm{cm^2 \, g^{-1}}$ to $0.1 \, \mathrm{cm^2 \, g^{-1}}$. However, when applying such a value, which is more typical of H- and He-poor Type Ic SNe \citep{Nagy2018ApJ}, to SN~2022ngb yields physically unrealistic values for the ejected mass and the explosion energy, supporting our choice of a higher opacity dominated by electron scattering.

The mass of synthesized $^{56}$Ni can be estimated from the diffusion time of the light curve, $\tau_{m}$, and its peak bolometric luminosity, $L_{\rm{Bol, peak}}$. According to Arnett's rule, it is approximated that the luminosity at the time of the peak ($t = \tau_{m}$) equals to the instantaneous rate of energy deposition from radioactive decay. This relationship is expressed as
\begin{equation}
L_{\rm{Bol}}(\tau_{m}) = L_{\rm{inp}}(\tau_{\rm{m}}) = M_{\rm{Ni}} [(\epsilon_{\rm{Ni}} - \epsilon_{\rm{Co}})e^{-\tau_{\rm{m}}/\tau_{\rm{Ni}}} + \epsilon_{\rm{Co}} e^{-\tau_{\rm{m}}/\tau_{\rm{Co}}}] \, .
\end{equation}
\noindent Here, $\epsilon_{Ni} = 3.9 \times 10^{10} \, \mathrm{erg\, s^{-1} g^{-1}}$ and $\epsilon_{Co} = 6.78 \times 10^{9} \, \mathrm{erg\, s^{-1} g^{-1}}$ are the specific energy generation rates for the decay of $^{56}$Ni and $^{56}$Co, respectively, while $\tau_{Ni} = 8.8 \, \mathrm{days}$ and $\tau_{Co} = 111.3 \, \mathrm{days}$ represent the characteristic decay timescales for these isotopes. Using the observed diffusion time, $\tau_{m} = 28.5\, \mathrm{days}$, and peak bolometric luminosity, $L_{\rm{Bol,\, peak}} = 7.76^{+1.15}_{-1.00} \times 10^{41} \, \mathrm{erg \, s^{-1}}$, this simple model yields a $^{56}$Ni mass for SN~2022ngb in the range from 0.05 \msun\ to 0.06 \msun. However, as suggested by \cite{Nagy2014AA}, energy released from hydrogen recombination can also contribute to the bolometric luminosity. This mechanics can be expressed as $L_{\mathrm{Bol}}(t)=L_{\mathrm{diff}}(t)+L_{\mathrm{rec}}(t)$, where the $L_{\mathrm{diff}}$ represents the diffusion luminosity mainly contribute by radioactive decay and $L_{\mathrm{rec}}$ represents the recombination contribution. Following the approaches provided by \mbox{\cite{Nagy2014AA}}, the $L_{\mathrm{rec}}$ can be expressed as
\begin{equation}
\mathrm{d}u_{\mathrm{rec}}=4\pi r_{\mathrm{rec}}^2 \rho(r,\,t) \epsilon_{\mathrm{rec}}\mathrm{d}r_{\mathrm{rec}};\;\;\;L_{\mathrm{rec}}=\dot u_{\mathrm{rec}}.\;\;
\end{equation}
\noindent Here, the $u_{\mathrm{rec}}$ is the energy contribute by recombination and $\epsilon_{\mathrm{rec}}$ is the recombination energy release per unit mass. The recombination front, $r_{\mathrm{rec}}$, is located at the radius, where $T(r,\,t)=T_{\mathrm{rec}}$.
Therefore, the value derived using the simple Arnett's rule should be considered an upper limit for the $^{56}$Ni mass. 

\begin{figure}[htbp]
\centering
\includegraphics[width=1.0\linewidth]{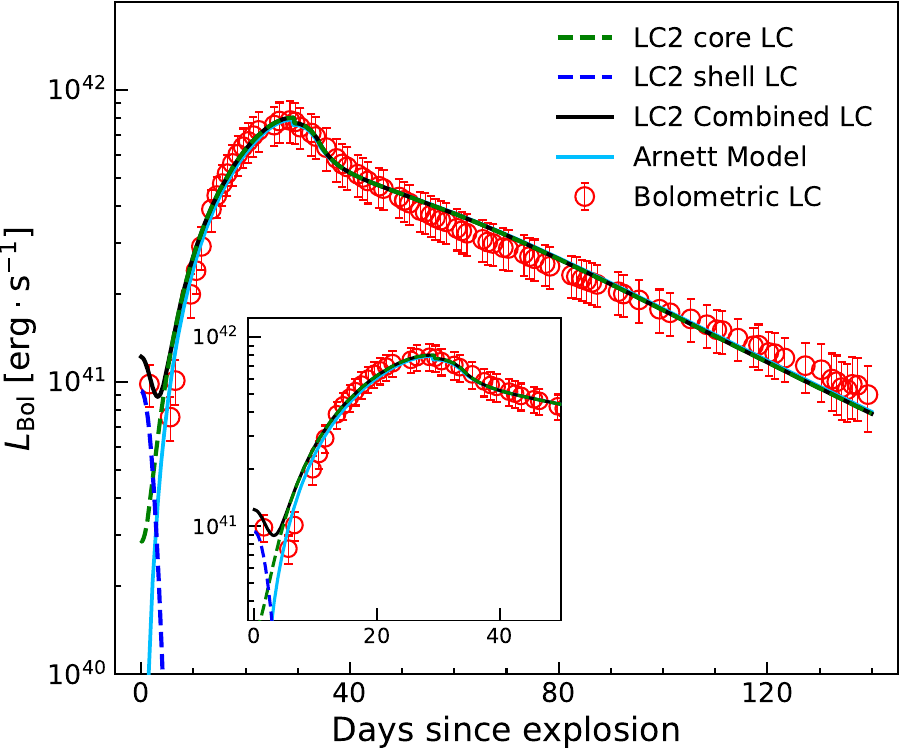}
\caption{Bolometric light curve fitting result for SN~2022ngb (red dots). The black solid line stands for the fitting result of the two-component model. Green and blue dashed lines stand for the contribution of the He-rich core and the H-rich shell, respectively. The cyan solid line is the result of the Arnett-like model fitting result. The inset shows the bolometric light curve of SN~2022ngb along with the fitting result up to $\sim 50$ days from explosion.
}
\label{bol_lc_fitting}
\end{figure}

To obtain a more precise estimate, we fit the bolometric light curve using a slightly modified Arnett's model with the contribution of recombination. This model improves upon the simple approximation by incorporating gamma-ray leakage and the energy contribution from recombination. For the fitting process, we adopted standard assumptions, including a small initial radius and negligible initial thermal energy. The best-fit model yields a $^{56}$Ni mass of 0.045 \msun, an ejecta mass of 3.2 \msun, and an initial kinetic energy of $1.34\times10^{51}\,\mathrm{erg}$. This refined $^{56}$Ni mass is, as expected, lower than the upper limit derived earlier. These physical parameters suggest a massive progenitor and an relatively energetic explosion, similar to SN~2020acat ($E_{\rm{k}}=1.4\times10^{51} \, \mathrm{erg}$; \citealt{Ergon2024A&A}). The resulting $^{56}$Ni mass is also comparable to that of SN~2024abfo ($M_{\rm{Ni}}\sim0.045\,\msun$; \citealt{Reguitti2025AA}), which exhibited a similar peak luminosity. The best-fit model is shown in Fig.~\ref{bol_lc_fitting}.

The two-component model from \cite{Nagy2016aap}, specifically developed for SNe IIb, is adopted to obtain a more precise estimate of the parameters of SN~2022ngb. This model consists of a dense, He-rich core, powered by $^{56}$Ni $\rightarrow$ $^{56}$Co decay, and an extended, low-mass, H-rich shell, powered by shock-cooling emission at early times. For SN~2022ngb, the early shock-cooling phase appears faint, with its contribution likely confined to the first few days ($\sim3$ days from explosion). Thus, the bolometric light curve is primarily dominated by the emission from the He-rich core. For the fitting process, we adopted fixed opacity values for each component. Based on its H-rich composition, the shell opacity was set to $\kappa_{sh} = 0.3$ cm$^2$ g$^{-1}$. For the He-rich core, we used $\kappa_{co} = 0.195$ cm$^2$ g$^{-1}$ following the previous estimation. Considering the parameters of our previous Arnett-like model, the resulting model and its parameters are presented in Fig.~\ref{bol_lc_fitting} and Table \ref{apptab_snlcfittingparams}, along with those of other SNe IIb whose parameters were inferred through the former Arnett-like model.

To estimate the mass and the radius of the hydrogen-rich envelope, we adopted the analytical model for shock-cooling emission described in \cite{Piro2021ApJ}. From the observed bolometric light curve, we noted that the initial cooling phase transitions to a radioactively powered rise at approximately +5 days after the explosion. Following the methodology of recent studies \citep[e.g.][]{Gangopadhyay2018mnras, Reguitti2025AA}, we used this transition time as an observational proxy for $t_{ph}$, the timescale for the photosphere to recede through the extended material. Therefore, we adopted $t_{ph} \approx 5 \, \mathrm{days}$. For the transition velocity $v_t$, we used the H$\alpha$ velocity of $19500 \, \mathrm{km \, s^{-1}}$ measured from our earliest available spectrum at $+$4.16 days from the explosion. We cautioned that this value should be considered a lower limit on the true transition velocity, as the photosphere had likely receded into slower moving ejecta by this epoch. For the opacity, we use a value typical for a hydrogen-rich shell, which ranges from $0.3 \, \mathrm{cm^2 \, g^{-1}}$ to $0.4 \, \mathrm{cm^2 \, g^{-1}}$ \citep{Nagy2016aap, Nagy2018ApJ}. Consequently, the envelope mass can be estimated using the following formula from \cite{Piro2021ApJ}:
\begin{equation}
M_{env} = \frac{2}{3} \frac{n-1}{\kappa K} t_{ph}^2 v_t^2, \;\; L(t) = \frac{\pi (n - 1) c v_{t}^2}{3 (n-5) \kappa} \left(\frac{t_d}{t}\right)^{4/(n-2)}R_e \, .
\end{equation}
Adopting the standard parameters $n = 10$ and $K = 0.119$ for the power-law density profile of the envelope \citep{Chevalier1989ApJ, Piro2021ApJ}, the mass of the hydrogen-rich envelope of SN~2022ngb was estimated as $\sim0.05 \, \msun$. We then estimated the initial envelope radius, $R_e$, using a shock-cooling model based on the early bolometric data. In this model, the emission is governed by the diffusion timescale of the envelope, $t_{\mathrm{d,\, env}} = \sqrt{3\kappa K M_{\mathrm{e}} / [(n - 1) v_{\mathrm{t}} c]}$. Applying this model to the observation at MJD 59751.5, when the SN luminosity was $L_{\mathrm{Bol}} = 9.84 \times 10^{40} \, \mathrm{erg \, s^{-1}}$, yielded an initial radius of $R_{\mathrm{e}}$ $\sim10^{11} \, \mathrm{cm}$, suggesting a relatively compact progenitor, similar to that of SN 2022crv. However, as noted by \cite{Reguitti2025AA}, radius estimates derived from the faint, early-stage shock cooling emission can be unreliable. Therefore, the value derived here should be considered an order-of-magnitude estimate. A more accurate determination would likely require a direct observation of the progenitor (not available in our dataset).

We present the detailed best-fit parameters for SN~2022ngb and other comparisons objects, as derived from the two-component and Arnett's approximation models in Table \ref{apptab_snlcfittingparams}. The core properties derived for SN~2022ngb are consistent with those of the comparison sample. In contrast, the properties of the outer shell differ significantly, suggesting a thinner, lower-mass residual hydrogen envelope (see Sect.  \ref{sec_shockcooling}).

To provide a complementary and self-consistent verification of the parameters derived from our semi-analytical modeling, we applied the scaling relations from \cite{Pumo2023MNRAS}. Adopting the methodology of \cite{Medler2021MNRAS} for reference selection, we utilized a set of well-characterized SESNe. Specifically, we selected the He-rich Type IIb and Ib events SN~1993J, SN~2003bg, and SN~2008D as primary references, and include the He-poor Type Ic events SN~2004aw and SN~2002ap for comparison. The scaling relations can be described as 
\begin{equation}
\begin{split}
E_{\mathrm{k}}&=E_{\mathrm{k,ref}}\times\left(\frac{\tau_{\mathrm{m}}}{\tau_{\mathrm{m,ref}}}\right)^2\times\left(\frac{v_{\mathrm{ej}}}{v_{\mathrm{ej,ref}}}\right)^3\times\left(\frac{\kappa_\mathrm{opt}}{\kappa_{\mathrm{opt,ref}}}\right)^{-1},\\
M_{\mathrm{ej}}&=M_{\mathrm{ej,ref}}\times\left(\frac{\tau_{\mathrm{m}}}{\tau_{\mathrm{m,ref}}}\right)^2\times\left(\frac{v_{\mathrm{ej}}}{v_{\mathrm{ej,ref}}}\right)\times\left(\frac{\kappa_\mathrm{opt}}{\kappa_{\mathrm{opt,ref}}}\right)^{-2}.\,
\end{split}
\end{equation}
Following \cite{Nagy2018ApJ}, we set the opacity in our modeling to $\kappa_{\mathrm{opt}}=0.195\,\mathrm{cm^2\,g^{-1}}$. We incorporated the effective opacity of the comparison objects derived by \cite{Medler2021MNRAS} directly into the scaling relations. This approach provides more robust constraints than simply assuming a same opacity across all objects. To mitigate uncertainties regarding the explosion epoch and rise time, we utilize the light curve width\footnote{defined as the width of the \textit{BgVri} pseudo-bolometric light curve at 0.5 magnitude below peak brightness}, measured here as $\sim24.31\,\mathrm{days}$. Applying these scaling relations using the He-rich reference scenarios yields $M_{\mathrm{ej}} =2.6\pm2.0\,\msun$ and $E_{\mathrm{k}}=(3.6\pm2.4)\times10^{51}\,\mathrm{erg}$ for SN~2022ngb. This result is consistent with our previous semi-analytical modeling, suggesting that our estimation is reasonable. A slight discrepancy arises in the kinetic energy, which is somewhat larger than the value derived from our modeling. We also examined the He-poor scenarios, which resulted in a lower ejecta mass ($\sim2.0\,\msun$) and a correspondingly lower kinetic energy ($\sim2.0\times10^{51}\,\mathrm{erg}$). These findings are fully consistent with the analysis presented by \cite{Medler2021MNRAS}, suggesting that SN~2022ngb follows the characteristics of typical Type~IIb events. 

As mentioned above, estimating the $^{56}$Ni mass through simple models may return quite uncertain values. For instance, \cite{Dessart2016MNRAS} note that when an Arnett-like model is applied to SESNe, the mass of $^{56}$Ni can be overestimated by as much as 40\% \citep[see, also,][]{Medler2021MNRAS, Medler2022MNRAS}, as the diffusion of $^{56}$Ni in the ejecta is neglected in Arnett-like models. In this case, the $^{56}$Ni mass of SN~2022ngb might be lower, of $\sim 0.03 \, \mathrm{M}_\odot$. To provide a more accurate result, we used SN~1987A as a reference, which has a well-constrained $^{56}$Ni mass of $0.075 \, \mathrm{M}_\odot$. According to \cite{Ravi2025apj}, the mass of $^{56}$Ni can be estimated with the following equation:
\begin{equation}
    M_{\rm{Ni}} = M_{\rm{Ni, 1987A}} \frac{1-\mathrm{exp}(-(530/t)^2)}{1-\mathrm{exp}(-(T_0/t)^2)} \frac{L(t)}{L_{\rm{1987A}}(t)} \, .
\end{equation}
\noindent Here, $T_0$ represents the timescale of gamma-ray leakage \citep[cf. 530~days for SN~1987A;][]{Jerkstrand-2011-PhDT}, which is crucial for the late-time light curve of SESNe \mbox{\citep{Clocchiatti1997ApJ}}. This returned a $^{56}$Ni mass of $0.035 \pm 0.008\, \mathrm{M}_\odot$, in good agreement with our previous estimate. Following \cite{Sukhbold2016apj}, a $^{56}$Ni mass of 0.04 \msun\ and an ejected mass of 3.0 \msun\ would favor an intermediate-mass progenitor, with an $\mathrm{M}_{ZAMS}$ around $15\, \, \mathrm{M}_\odot$, hence generating a compact remnant with a mass of about $1.5 \, \mathrm{M}_\odot$. Thus, the progenitor of SN~2022ngb is likely similar to those of SN~2020acat ($\sim 17\,\msun$; \citealt{Ergon2024A&A}) and SN~2008ax ($\sim 18 \, \mathrm{M}_\odot$; \citealt{Folatelli2015apj}).

\label{sec:lc_model}
\if
1993J: \citet{Richmond1994aj}, 2. \citet{Barbon1995aaps}, 3. \citet{Richmond1996aj}
2008ax: \citet{Pastorello2008mnras}, 
2011dh: \citet{Sahu2013mnras},\citet{Ergon2015aap}
2011fu: \citet{Kumar2013mnras},\citet{Morales-Garoffolo2015mnras}
2013df: \citet{Morales-Garoffolo2014mnras},\citet{VanDyk2014aj}
2015as: \citet{Gangopadhyay2018mnras}
2016gkg: \citet{Tartaglia2017apjl},\citet{Arcavi2017apjl},\citet{Bersten2018nat}
2020acat: \citet{Medler2022MNRAS},\citet{Ergon2024A&A}
2021bxu: \citet{Desai2023mnras}
\fi

\begin{figure*}[htbp]
\centering
\includegraphics[width=1.0\linewidth]{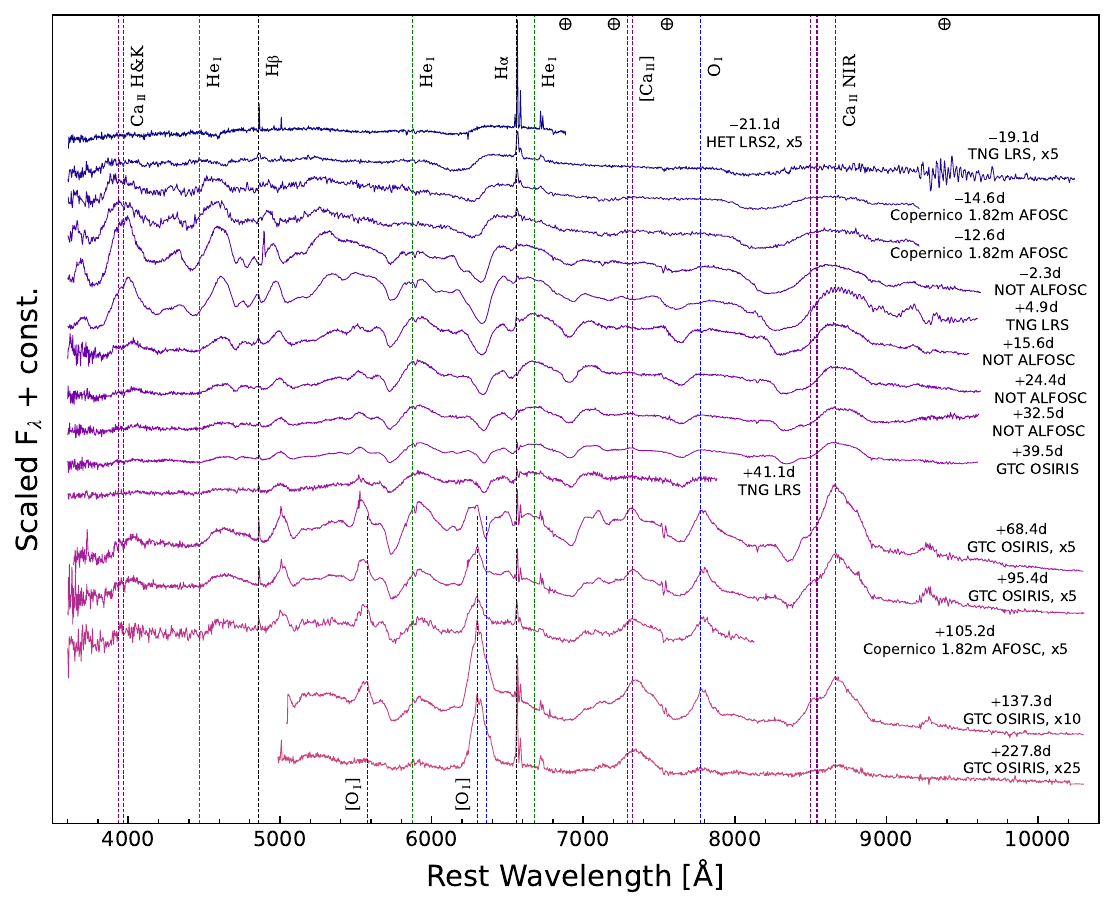}
\caption{Spectroscopic evolution of SN~2022ngb, with prominent spectral features marked with vertical lines. The phase (from \textit{V}-band maximum) of each spectrum and associated instruments are indicated on the right. The major telluric bands are marked with a telluric symbol at the top.}
\label{spectrum}
\end{figure*}
\begin{table*}[htbp]
\centering
\caption{Spectroscopic observations of SN\,2022ngb.}
\label{tab_2022ngbSpecInfo} 
\begin{tabular}{@{\extracolsep{0.1em}}cccccccc} \\
\hline \hline
Date & MJD & Phase$^a$ & Telescope+Instrument & Grism/Grating+Slit & Spectral range & Resolution & Exp. time \\ 
  &   & (days) &   &        & (\AA)    & (\AA)           & (s)           \\ 
\hline
20220623 & 59753.4 & $-$21.1 & HET + LRS2 & UV+orange & 3640-6950 & R=1910;R=1140 & 1800 \\
20220624 & 59755.4 & $-$19.1 & TNG + LRS  & LRB+LRR+1.5" & 3260-10340 & 14;15 & 1500 \\
20220629 & 59760.0 & $-$14.6 & Copernico + AFOSC & VPH6+VPH7+1.69" & 3160-9300 & 14 & 2400 \\
20220701 & 59761.9 & $-$12.6 & Copernico + AFOSC & VPH6+VPH7+1.69" & 3160-9300 & 14 & 2400 \\
20220712 & 59772.2 & $-$2.3 & NOT + ALFOSC & gm4+1.0" & 3400-9710 & 14 & 1800 \\
20220718 & 59779.4 & +4.9 & TNG + LRS & LRB+LRR+1.5" & 3260-10350 & 13.5;15 & 2100 \\
20220730 & 59790.1 & +15.6 & NOT + ALFOSC & gm4+1.3" & 3400-9630 & 18 & 1800 \\
20220807 & 59799.0 & +24.4 & NOT + ALFOSC & gm4+1.0" & 3400-9710 & 14 & 1800 \\
20220816 & 59807.0 & +32.5 & NOT + ALFOSC & gm4+1.0" & 3430-9700 & 14 & 1800 \\
20220822 & 59814.0 & +39.5 & GTC + OSIRIS & R1000B+R1000R+1.0" & 3630-10400 & 7;12  & 900 \\
20220824 & 59815.6 & +41.1 & TNG + LRS & LR-B+1.0" & 3260-7960 & 11 & 1800 \\
20220920 & 59843.0 & +68.4 & GTC + OSIRIS & R1000B+R1000R+1.0" & 3670-10400 & 7;12 & 1200 \\
20221017 & 59869.9 & +95.4 & GTC + OSIRIS & R1000B+R1000R+1.0" & 3630-10400 & 7;13 & 1600 \\
20221027 & 59879.7 & +105.2 & Copernico + AFOSC & GR04+1.69" & 3410-8200 & 13 & 3600 \\
20221128 & 59911.8 & +137.3 & GTC + OSIRIS & R1000B+R1000R+1.0" & 5090-10400 & 8;13 & 1350 \\
20230227 & 60002.3 & +227.8 & GTC + OSIRIS & R1000B+R1000R+1.0" & 5030-10400 & 8;13 & 1000 \\
\hline
\multicolumn{7}{l}{{$^a$Phases are relative to $V$-band maximum light.  }} \\
\end{tabular}
\end{table*}
\section{Spectroscopy}
\label{sec_spec}
\subsection{Spectral evolution}
We present the full set of spectra of SN~2022ngb in Fig.~\ref{spectrum}. The log of the spectroscopic observations is provided in Table~\ref{tab_2022ngbSpecInfo}. All spectra were flux-calibrated with the photometric data and corrected for MW and host galaxy reddening and redshift. The spectra cover a period from 21.1 days before to 227.8 days after the \textit{V}-band maximum, with the latest data extending into the nebular phase. Narrow emission lines unresolved, such as \Ha, \Hb, \Nii $\mathrm{\lambda \lambda}$6548,6583~\AA, and \Sii $\mathrm{\lambda \lambda}$6716,6731~\AA, originate from \Hii regions in the host galaxy. 

In the early phases ($-$21.1, $-$19.1, and $-$14.6 days from \textit{V}-band maximum), the SN exhibits a relatively blue continuum. Throughout its evolution, the continuum gradually becomes redder, until prominent emission lines show up in the late-phase spectra ($+$68.4 days since \textit{V}-band maximum and later), marking the transition to the nebular phase.

In the first two spectra, we noticed several broad lines with a P~Cygni profile, such as \Ha, \Hb, \Siliconii, and a barely detectable \Hei~5876~\AA~feature (see Sect. \ref{sec_synapps}, for more details). Starting from around $-$14 to $-$12 days, several broad metal lines appear on the blue side of the spectrum, exhibiting strong P~Cygni profiles indicative of fast-expanding ejecta. From $-$14.6 days onward, the blue region of the spectra is dominated by \Feii, with blended absorption lines of \Scii and \Tiii. Additionally, a blue-shifted \Caiinforb H\&K feature is visible at around 3800 \AA. As the SN evolves, the \Hei lines become more prominent, while the \Ha line begins to fade after $+$32.5 d. This behavior is characteristic of the spectral evolution of a SN IIb, suggesting the presence of a thin hydrogen layer overlying a He-rich core.

A transitional phase begins at around $+$32.5 days post \textit{V}-band maximum, as the outer ejecta become increasingly transparent and prominent emission lines emerge. As the photosphere retreats, the Balmer lines progressively weaken and become narrower. The \Ha line, in particular, diminishes while \Hei absorption becomes dominant from +68.4 days onward, causing the spectra to evolve into a Type Ib-like appearance. This evolutionary path is consistent with observations of other SNe IIb \citep[see, e.g.][]{Medler2022MNRAS}. With evolution, strong forbidden emission lines from \Oi and \Caii emerge. Additionally, the \Caiinforb NIR triplet appears and strengthens significantly during this period. These late-time spectral features, typical of the nebular phase of a CCSN, probe the products synthesized in the core and provide crucial constraints on the structure of the progenitor.

\subsection{\texttt{SYNAPPS} modeling}
\label{sec_synapps}
\begin{figure*}[htbp]
\centering
\includegraphics[width=1.0\linewidth]{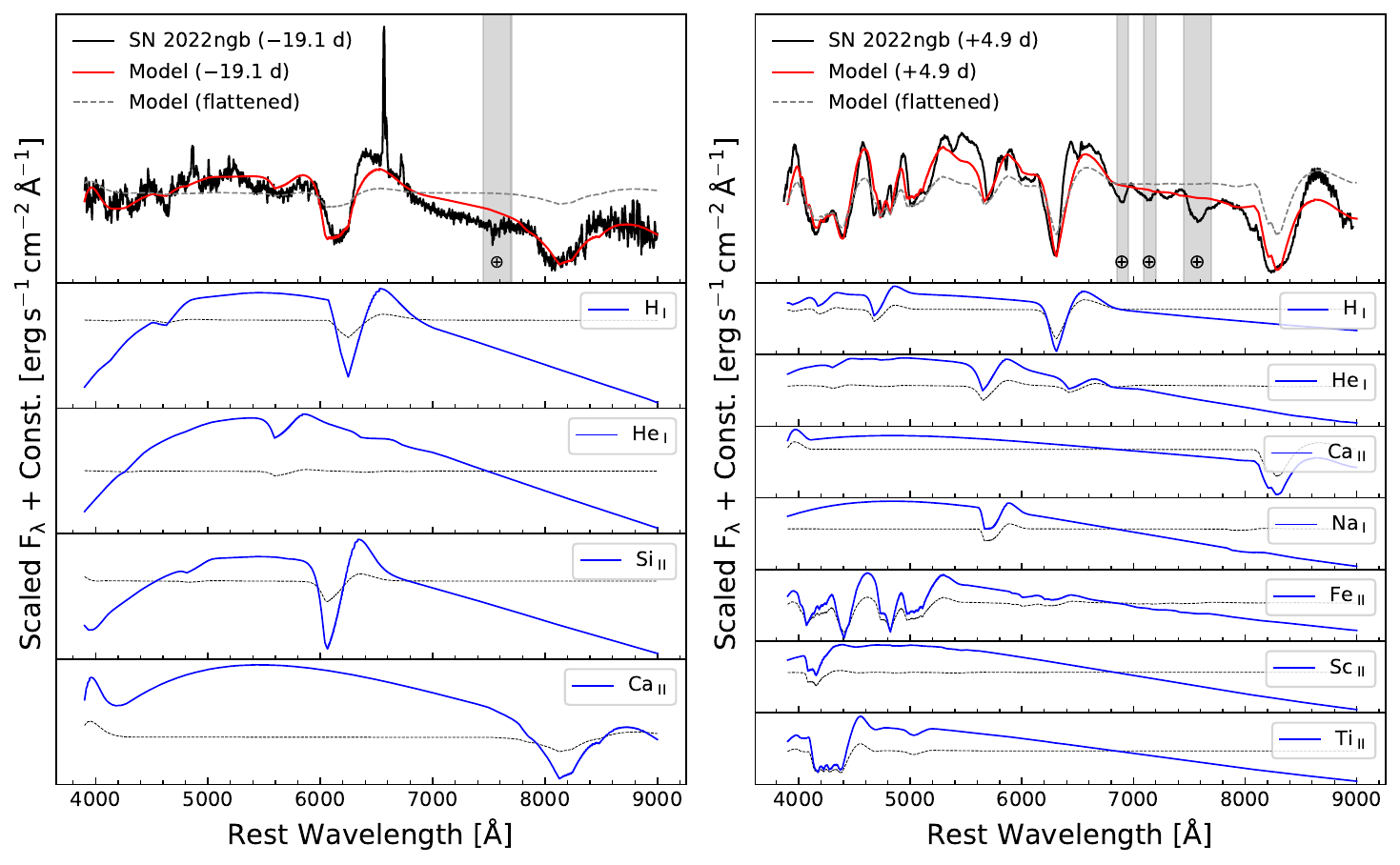}
\caption{Early ($-$19.1 days from \textit{V}-band maximum; top-left panel) and near maximum ($+$4.9 days from \textit{V}-band maximum; top-right panel) spectra of SN~2022ngb compared with spectral models generated by \texttt{SYN++}. Solid lines represent the best-fit models, while the dashed lines present the continuum-subtracted result. The most important contributing species are shown in the lower panels.
}
\label{spec_app}
\end{figure*}

To ensure the identification of spectral lines, we modeled an early-phase spectrum ($-$19.1 days from \textit{V}-band maximum) and a spectrum around the peak ($+$4.9 days from \textit{V}-band maximum). Following the procedure for handling spectral identifications in SNe IIb \cite[e.g.][]{Gangopadhyay2023ApJ, zou2025arXiv}, we use \texttt{SYN++} \citep{Thomas2013ascl} in conjunction with \texttt{SYNAPPS} \citep{Thomas2011pasp} to create a data-driven model for our spectra, to support the identification of significant spectral features. In this model, all photons are assumed to escape from the photosphere, a region defined by the optical depth as a function of velocity. Our results are presented in Fig.~\ref{spec_app}. The main panel displays the observed data and the best-fit model (with and without continuum-subtracted), while the subplots show the individual line components constructed by \texttt{SYN++} using the best-fit parameters. Each of these spectral lines exhibits a prominent P~Cygni profile, and the overall model well fits our observational data.

To eliminate the influence of telluric absorption and narrow emission lines from the host galaxy \Hii\ regions, we simply excluded these spectral regions from the fitting process, as indicated in Fig.~\ref{spec_app}. In the early-phase spectrum, we identify prominent features of H, \Siliconii, and \Caiinforb. Although \Nai\ and \Feii\ lines were also taken into account, their contribution appears to be negligible, and are not shown in the figure. We notice that all Balmer lines exhibit broad and significantly blue-shifted profiles, suggesting rapidly expanding ejecta at a high temperature. In this model, the best-fit photospheric velocity is 15300 $\mathrm{km \, s^{-1}}$ and the temperature is 6500 K. This temperature is consistent with values derived from blackbody fits to both the bolometric light curve and the spectral continuum and is also 
comparable to that of other single-peaked SNe IIb \citep[e.g., SN~2015as, which had a temperature of around 8000 K at a similar epoch,][]{Gangopadhyay2018mnras}. 
In contrast, SNe IIb with prominent double-peaked light curves (e.g., SN~1993J, SN~2024aecx) typically have higher temperatures. For instance, the photospheric temperature of SN~2024aecx at a similar stage was around 14000 K \citep{zou2025arXiv}.

During the early phase, we observed a broad absorption feature near 6200 \AA\, mainly corresponding to the P~Cygni profile of \Ha\ (Fig.~\ref{spec_app}). In contrast, the \Hb\ line does not exhibit a similarly broad profile. This disparity suggests that the 6200~\AA\ feature is not produced solely by hydrogen, a phenomenon previously observed in other SNe IIb. We explored several possible explanations for this feature. Although some studies suggest this feature in SESNe is a blend of \Ha, \Siliconii, \Feii, and \Cii\ \citep[e.g.][]{Elmhamdi2006AA, Holmbo2023A&A}, our spectral fitting indicates that contributions from \Feii\ and \Cii\ are negligible in this case. Another possibility, a double-velocity component in the \Ha\ line as proposed by \cite{Milisavljevic2013apj} and \cite{Medler2021MNRAS}, is also unlikely. We found no evidence of two distinct absorption components in our spectrum, unlike the early spectra of SN~2011ei and SN~2020cpg. Instead, the feature in our spectrum resembles that observed in SN~2015as, for which \cite{Gangopadhyay2018mnras} attributed the broad absorption to a combined contribution from \Ha\ and \Siliconii. We adopted this interpretation and our synthetic spectrum provides a good fit to the data, which is also consistent with the template spectra provided by \cite{Holmbo2023A&A}. Our early-phase spectral model also reproduces other observed features. We identified a faint P~Cygni profile of \Hei\ line near 5750~\AA\ (left panel of Fig.~\ref{spec_app}), and this feature becomes prominent around $-$12.6 days and $-$2.3 days from \textit{V}-band maximum, which follows the typical spectral evolution of Type~IIb SNe. Furthermore, in the NIR region, our model reproduces a prominent absorption feature from the \Caiinforb\ triplet at wavelengths greater than 8000~\AA.

We also present the analysis of a near-maximum ($+$4.9 days) spectrum, with the best-fit model shown in the right panel of Fig.~\ref{spec_app}. On the blue side of the spectrum, below 5500 \AA, we identified strong \Feii features blended with \Scii and \Tiii. The presence of these features indicates a decrease in temperature, which allows for the formation of singly ionized metal ions. Furthermore, the \Ha line has become narrower compared to earlier epochs, while the \Hei absorption lines show a growing trend. The blue side of the spectrum is dominated by metal lines, particularly \Feii and the \Caiinforb H\&K (shown in Fig.~\ref{spectrum}), both of which exhibit clear P~Cygni profiles. Furthermore, we found a strong absorption feature around 5800~\AA, which our fitting result attributes to a combined contribution from both \Nai and \Hei. We also noted that a \Hei contribution is still present on the blue side of the feature, although blended with metal lines.

\subsection{Spectral comparison with other SNe IIb}
\label{speccmp_sec}
\begin{figure*}[htbp]
\centering
\includegraphics[width=1.0\linewidth]{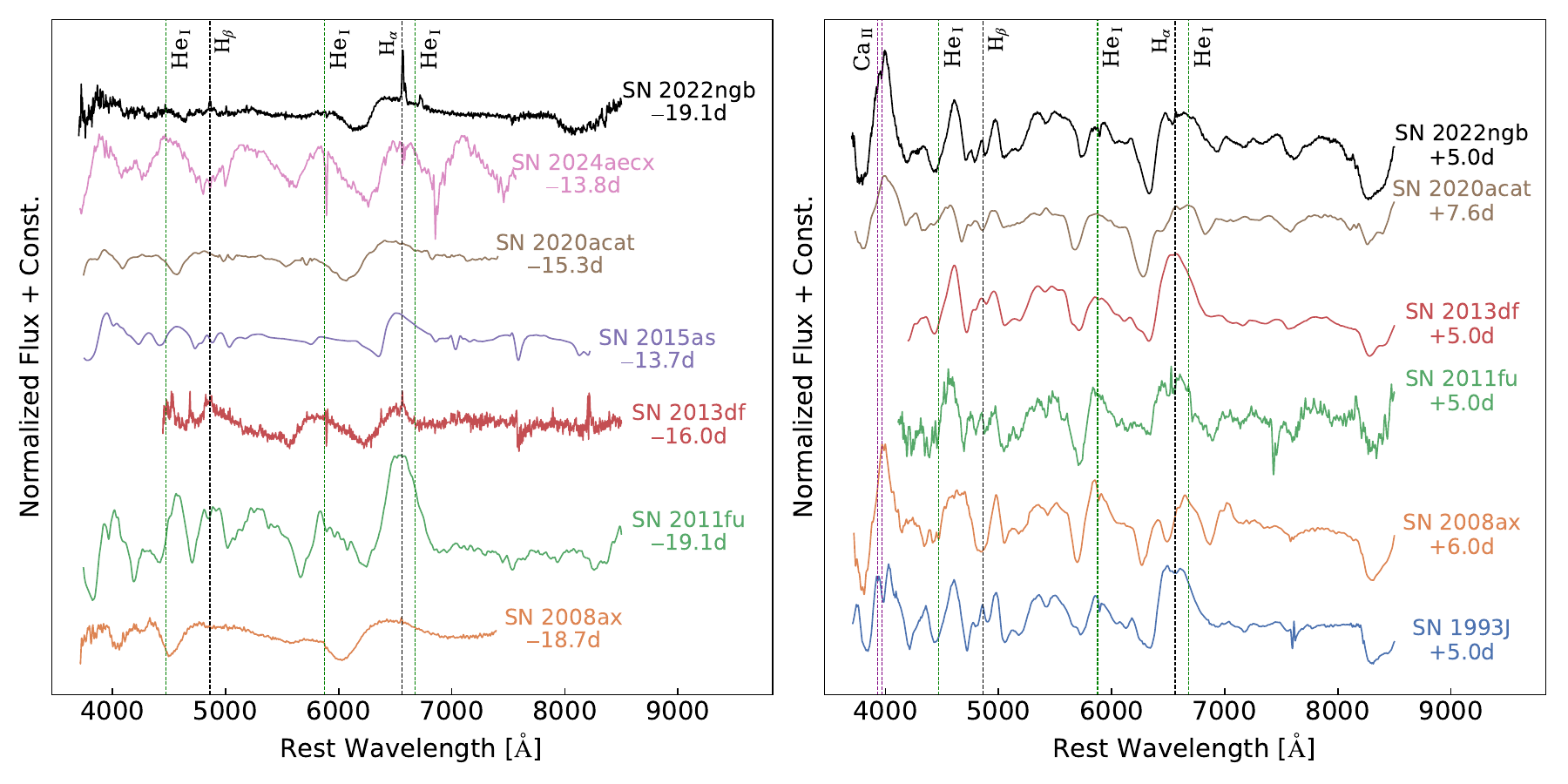}
\caption{Comparisons of the spectra of SN~2022ngb with other Type IIb events (SN~1993J, SN~2008ax, SN~2011fu, SN~2013df, SN~2015as, SN~2020acat, and SN~2024aecx) taken before (left panel) and soon after the maximum light (right panel). Key spectral lines are highlighted with corresponding labels. The names of the comparison SNe and their respective phases are displayed on the right side of each spectrum.
}
\label{spec_cmp}
\end{figure*}

\begin{figure}[htbp]
\centering
\includegraphics[width=1.0\linewidth]{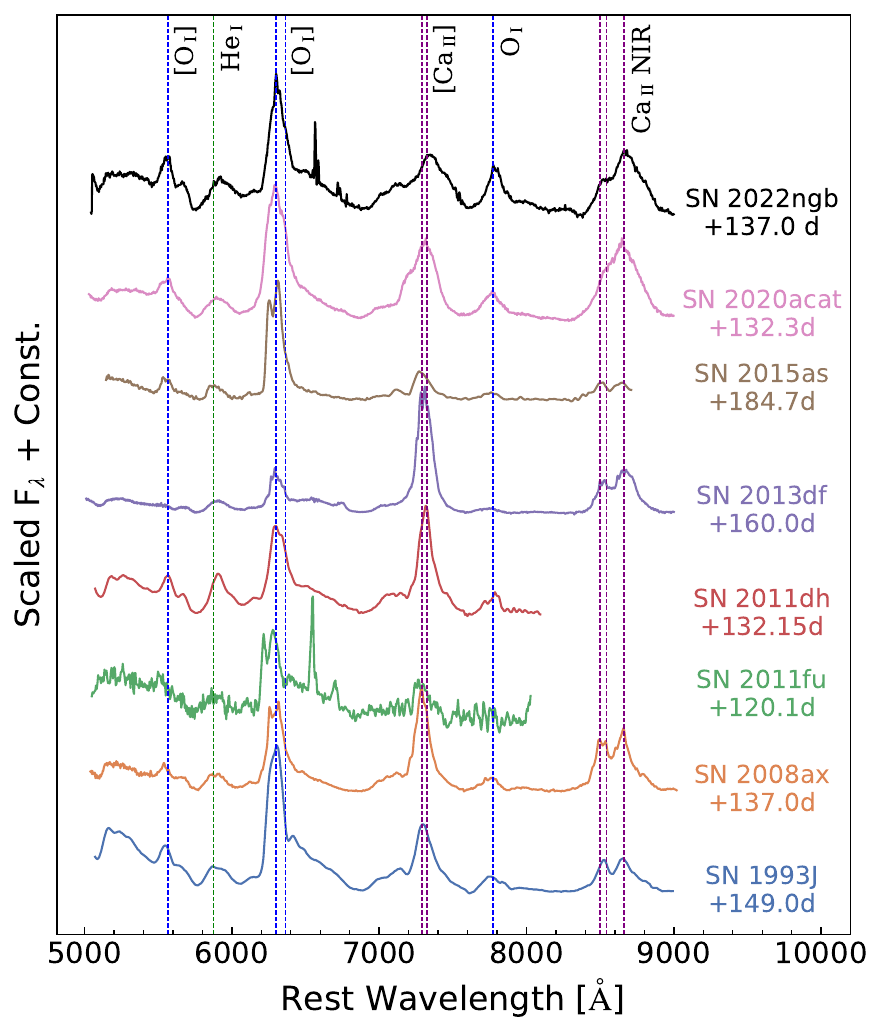}
\caption{Comparison of the nebular spectrum of SN~2022ngb with those of other Type IIb events (SN~1993J, SN~2008ax, SN~2011fu, SN~2013df, SN~2015as, and SN~2020acat) a Significant spectral lines are highlighted with corresponding labels. The names of the comparison SNe and their respective phases are displayed on the right side of each spectrum.
}
\label{spec_cmp_neb}
\end{figure}

To better characterize SN~2022ngb in the context of typical SN IIb, we compared its spectra with those of well-studied objects at similar phases\footnote{The spectra of the comparison objects were retrieved from the WISeRep database \citep{Yaron2012pasp}.} Individual spectra were corrected for redshift and line-of-sight extinction. In particular, Fig. \ref{spec_cmp} shows a spectral comparison of SN~2022ngb with other SNe IIb at a very early phase (2-3 weeks before the maximum light; left panel) and soon after peak (5-7 days after maximum; right panel). Figure~\ref{spec_cmp_neb} shows a comparison in the nebular phase (4-6 months after maximum). All epochs were calculated relative to the \textit{V}-band maximum. Significant spectral lines are identified and marked in each figure. For a clearer comparison, the early and near-maximum spectra were normalized using a continuum fit. The nebular phase spectra, were scaled by a constant factor to better highlight key features.
The pre-maximum spectral evolution of SN~2022ngb, presented in the left panels of Fig.~\ref{spec_cmp}, reveals several key properties of the explosion. The blueshift of the P~Cygni absorption feature around 6200~\AA, which is mainly contributed by \Ha and may also be blended with \Siliconii, indicates a high expansion velocity. The velocity of the ejecta, which infer from possible P~Cygni absorption feature of \Ha and \Hb lines in the early phase is comparable to that of SN~2013df \citep{Morales-Garoffolo2014mnras} and SN~2024aecx \citep{zou2025arXiv}, although lower than that of the more energetic SN~2020acat and SN~2008ax. A notable characteristic in the early spectra is the broad P Cygni absorption near 6200~\AA. As discussed in the previous section, the flat bottom profile of this feature suggests a blending between the \Ha\ and \Siliconii\ $\mathrm{\lambda}$6355 P Cygni absorption lines. Furthermore, the \Caiinforb\ NIR triplet is present as a prominent and broad absorption feature during this phase. In contrast, the \Hei\ features appear significantly weaker than those observed in other SNe IIb at comparable epochs.

Taken together, considering our bolometric light curve fitting result and our previous analysis, we conclude that the ejecta of SN~2022ngb are denser than those of other SNe IIb. In such a scenario, the high density would result in a higher optical depth, keeping the photosphere in the outer layers. This would naturally mask the features of the underlying helium layer. Additionally, we observed strong \Siliconii and \Caiinforb NIR absorption features in the early-time spectra of SN~2022ngb, which are more pronounced than in other SNe IIb. The velocity of \Siliconii $\mathrm{\lambda}$6355 is around 13500 $\mathrm{km \, s^{-1}}$, while the \Caiinforb NIR triplet has a velocity of $\sim$16000 $\mathrm{km \, s^{-1}}$ (both velocities result from the best-fit \texttt{SYNAPPS} model). 

\cite{Dessart2016MNRAS} suggested that mixing phenomena commonly occur in Type IIb events and that an appropriate degree of mixing enables an event to evolve along the canonical Type IIb pathway. The strength of this mixing is a key factor. \cite{Jerkstrand2015aap} suggested that mixing can push radioactive clumps toward the surface of the ejecta, which in turn affects the light curve and spectral shape. However, for SN~2022ngb, we did not find the features typical of the strong mixing events described in these articles. As noted in our previous light curve analysis, SN~2022ngb has a slower rise rate,  a low peak luminosity of about $L_{\mathrm{Bol}} = 7.76^{+1.15}_{-1.00} \times 10^{41} \, \mathrm{erg \, s^{-1}}$, and a longer light-curve rise time of around 28.5 days. 

\cite{Baal2024MNRAS} performed 3D hydrodynamic simulations with an NLTE radiative transfer code. This research indicates that silicon and calcium are formed during the explosion, and their early appearance is a signature of an asymmetric explosion. In this scenario, high-velocity silicon and calcium are injected locally into the outer ejecta because of the asymmetric explosion and hydrodynamic instabilities. \cite{Baal2024MNRAS} also noted that in less energetic Type IIb events, this local mixing can be more obvious. This intrinsic explosion asymmetry suggests that other elements, such as nickel, might follow a different and stronger macroscopic mixing behavior. Taking this into account, SN~2022ngb, with its lower explosion energy ($1.32\times10^{51}\,\mathrm{erg}$) than SN~1993J, SN~2011fu (both $\sim2.4\times10^{51}\,\mathrm{erg}$; \citealt{Nagy2016aap}) and comparable to SN~2020acat ($\sim1.2\times10^{51}\,\mathrm{erg}$; \citealt{Medler2022MNRAS, Ergon2024A&A}), can be aptly explained by this scenario. This suggests that SN~2022ngb was the result of an asymmetric explosion, which in some cases can spectroscopically produce features that mimic those of an intermediate-mixing event. In addition, this kind of asymmetry might not mix material from the inner core into the Hydrogen-rich shell along all sight lines, which is also consistent with our observational data.

The right panel of Fig.~\ref{spec_cmp} shows the spectra around the \textit{V}-band maximum for each event. On the bluer side of each spectrum, absorption caused by \Caiinforb, \Feii, \Scii, and \Tiii is prominent. All spectra show a strong P Cygni profile, presenting the typical features of an expanding ejecta with a well-defined photosphere. As the photosphere retreats and the ejecta becomes progressively more transparent, the \Hei lines become prominent in each spectrum. Compared with SNe 1993J, 2011fu, and 2013df, the \Hei $\mathrm{\lambda}$5875 \AA\ line in SN~2022ngb exhibits a sharp and deep absorption component. This suggests that the He is located in a well-separated shell. Furthermore, considering the evolution of \Ha, its line profile becomes narrower compared to the very early phase. We also note another absorption feature on the blue side of the \Ha profile. We propose that the absorption feature of the P Cygni profile around 6200 \AA\ is most likely a combination of \Ha\ and \Siliconii, consistent with the interpretation of \citet{Gangopadhyay2018mnras} and our previous modeling of the early spectrum (see the best-fit result in Sect. \ref{sec_synapps}). This will be further discussed in Sect.~\ref{sec_haevo}.

A comparison of nebular phase spectra of SN~2022ngb and other SNe IIb is presented in Fig.~\ref{spec_cmp_neb}. During this phase, the density of ejecta decreases and the photosphere recedes, enabling the analysis of the inner ejecta. We observed that the \Ha\ line typically disappears or becomes very weak, while the \Hei lines decrease in strength. At the same time, the \Oi $\mathrm{\lambda \lambda}$6300,6364 \AA\ and \Caii $\mathrm{\lambda \lambda}$7291,7323 \AA\ increase in strength, eventually dominating the optical spectrum. In the bluer region of the spectrum, the \Oi $\mathrm{\lambda}$5577 \AA\ line is present in SN~2022ngb and most SNe IIb, with a notable difference between SN~2013df, which exhibits only a weak oxygen emission feature. The presence of prominent helium lines alongside a weak or absent hydrogen line is a defining characteristic of SNe IIb, signifying their evolutionary transition between Type II and Type Ib. Furthermore, we identified the \Oinforb $\mathrm{\lambda \lambda}$7772,7774 \AA\ and \Caiinforb NIR triplet lines in SN~2022ngb and the comparison objects. We noted that the \Oinforb $\mathrm{\lambda \lambda}$7772,7774 \AA\ emission in SN~2022ngb appears to be stronger than in the comparison sample, which may indicate a more massive oxygen component in its ejecta.

Although the SNe IIb in our comparison sample share similar spectral features, they exhibit distinct differences in their profiles and specific line characteristics. In SN~2022ngb, the oxygen emission lines appear to be systematically stronger and display multipeaked profiles. While both SN~2008ax \citep{Pastorello2008mnras} and SN~2011dh \citep{Sahu2013mnras} also show a double-peaked profile in the \Oi $\mathrm{\lambda \lambda}$6300, 6364 \AA\ doublet, the peak separation of the doublet components in SN~2022ngb is only $\sim30$ \AA, a factor of two narrower than the comparison objects The \Caii $\mathrm{\lambda \lambda}$7291, 7323 feature in SN~2022ngb is relatively broad compared to the other SNe, with the notable exception of SN~2020acat. This broadness is also apparent in the permitted \Caiinforb NIR triplet, which is again consistent with SN~2020acat and would indicate that for both SNe, there is a wide velocity distribution for the calcium-rich ejecta. Finally, the rounded peak profile of the \Caii feature in SN~2022ngb shows a similarity to that of SN~2015as \citep{Gangopadhyay2018mnras}.

We also noted the presence of subtle emission features on the red wing of the \Oi $\mathrm{\lambda \lambda}$6300, 6364 \AA\ doublet in SN~2022ngb, similar to those observed in SN~1993J and SN~2011fu, interpreted as evidence of late-time interaction with hydrogen-rich CSM \citep{Maurer2010MNRAS, Sahu2013mnras}. However, confirming a CSM interaction scenario for SN~2022ngb is challenging. We lack late-time NIR spectroscopy that could reveal interaction, and we also did not observe the box-shaped \Ha\ emission lines at later phases, as suggested by \cite{deWet2025A&A...704A..89D}. Given the lack of strong evidence to support CSM interaction, we explored an alternative interpretation. \cite{Jerkstrand2015aap} suggested that in SNe IIb at this phase (around 150 days post-explosion, $+$162 days in our case), the contributions from hydrogen and helium at the red wing of \Oi $\mathrm{\lambda \lambda} 6300,\,6364$\AA\ are expected to be negligible. This drove us to consider the contribution of \Nii $\mathrm{\lambda \lambda}6548, 6583$ \AA, which \cite{Barmentloo2024MNRAS} proposed as the origin of the structure on the red wing of the \Oi doublet. According to their work, the strength of this \Nii feature is inversely correlated with the mass of the progenitor, making it a potential probe of the progenitor. As a consequence, the moderate strength of this feature in SN~2022ngb allows us to qualitatively estimate its progenitor mass to be likely lower than that of SN~2020acat (which exhibits a negligible \Nii feature, corresponding to a $M_{\rm{ZAMS}}\approx17\,\msun$, \citealt{Ergon2024A&A}), but higher than that of SN~2011dh (with prominent \Nii profiles).

\subsection{Line velocities}
\label{linevelocity}

\begin{figure}[htbp]
\centering
\includegraphics[width=1.0\linewidth]{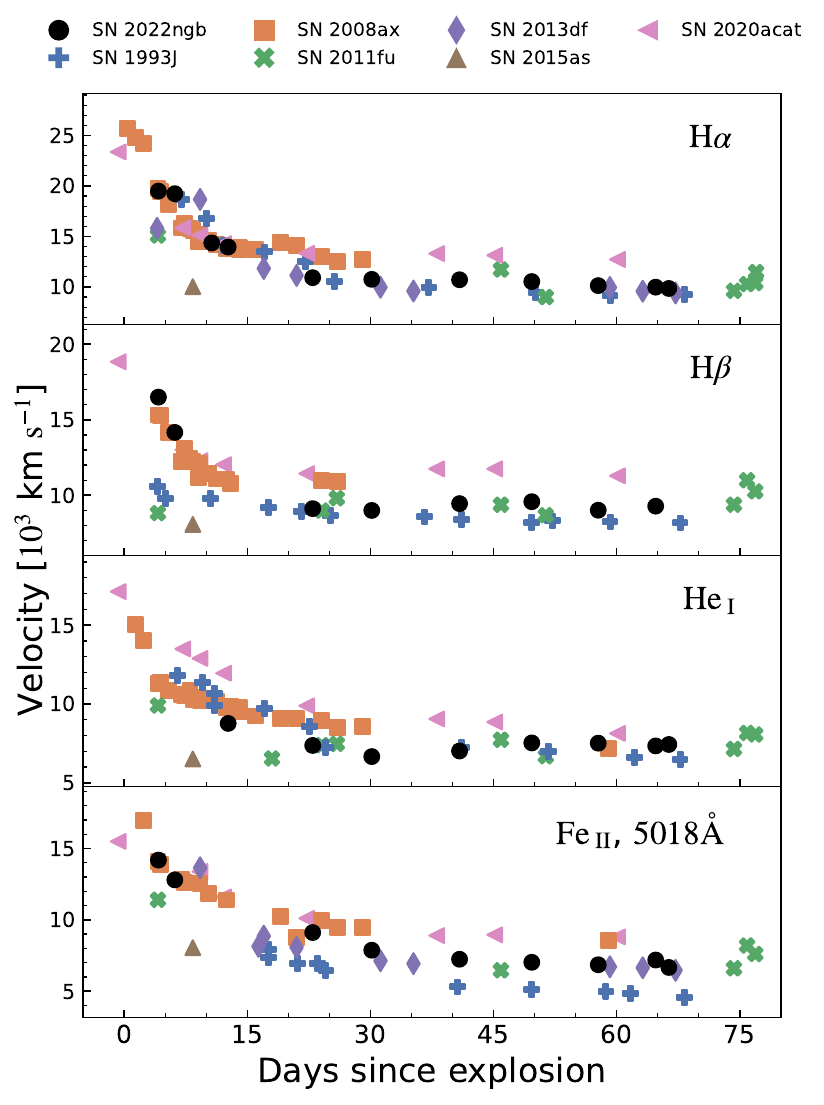}
\caption{Evolution of the line velocities for \Ha, \Hb, \Hei, and \Feii. SN~2022ngb and other comparisons are marked with different marker shape. The measurement errors are not shown here, but they can be up to 800 $\mathrm{km \, s^{-1}}$.
}
\label{spec_velocity}
\end{figure}

We performed Gaussian profile fits to the major absorption troughs in the spectra of SN~2022ngb to measure their velocities. Our analysis focuses on the \Ha\ $\mathrm{\lambda}6563$ \AA, \Hb\ $\mathrm{\lambda}4861$ \AA, \Hei\ $\mathrm{\lambda}5876$ \AA, and \Feii\ $\mathrm{\lambda}5018$ \AA\ lines. We then compared the velocity evolution of SN~2022ngb with those of other SNe IIb. The evolution of the line velocities in SN~2022ngb and the comparison sample is presented in Fig.~\ref{spec_velocity}. The typical uncertainty in our velocity measurements is up to 800 $\mathrm{km \, s^{-1}}$. For the \Feii\ line at early phases, however, the uncertainty is as high as 1000 $\mathrm{km \, s^{-1}}$ due to the lower S/N spectra and the weakeness of the Fe II spectral features.

The velocity evolution of SN~2022ngb is generally consistent with that of most SNe IIb in our comparison sample. Specifically, its velocity is higher than that of SN~2011fu and SN~2015as, and similar to higher-energy events such as SN~2020acat. In the very early stage, SN~2022ngb displayed high expansion velocities, with \Ha and \Hb line velocities of 19600 $\mathrm{km \, s^{-1}}$ and 16600 $\mathrm{km \, s^{-1}}$, respectively. We also identified a weak \Feii feature in our earliest spectra, which enabled us to provide an approximate velocity measurement. We found that the \Feii velocity during the early phase (up to 15 days from explosion) was around 15000 $\mathrm{km \, s^{-1}}$. This value is consistent with the best-fit photospheric velocity derived from \texttt{SYNAPPS} modeling, suggesting that the photosphere was located near the outer edge of the dense, opaque ejecta at this time. During the transition phase (from 20 to 70 days after the explosion), the \Hei line velocity in SN~2022ngb was systematically lower than that of \Ha, a behavior consistent with those of most SNe IIb \citep{Liu2016ApJ}. In this period, the velocity of \Feii\ become slower than the earlier phases. The \Feii feature also became more prominent, reducing the uncertainties in the velocity measurements. In the late transition phase, the emergence of the \Oi $\mathrm{\lambda \lambda}$6300, 6364 \AA\ doublet near the \Ha line profile prevented us from carrying out a reliable measurement of the \Ha velocity.

The evolution of velocity reveals a distinct layered structure within the ejecta of SN~2022ngb, a phenomenon also observed in SN~2015as \citep{Gangopadhyay2018mnras}. This structure comprises an outer layer of hydrogen expanding initially at a high velocity, and decelerating from approximately 20000 $\mathrm{km \, s^{-1}}$ at early times to 12000 $\mathrm{km \, s^{-1}}$ later. Beneath this lies an inner layer containing a mixture of iron and helium, which expands more slowly, with its velocity decreasing from about 15000 $\mathrm{km \, s^{-1}}$ to 6000 $\mathrm{km \, s^{-1}}$ over the same period. This velocity stratification reminds that found for SN~1993J \citep{Shigeyama1994ApJ}. In that case, it was proposed that Rayleigh-Taylor instabilities were inefficient at inducing large-scale mixing in the compact hydrogen envelopes typical of Type IIb events. Our observation of relatively clean absorption profiles, albeit with some features potentially indicative of mixing due to a localized, asymmetric explosion, corroborates this interpretation. 

Theoretical models proposed by \cite{Iwamoto1997ApJ} establish a relationship between the expansion velocity of the hydrogen envelope, the explosion energy $E_{\rm{exp}}$, and the ejecta mass $M_{\rm{ej}}$, described by the scaling relation $v_{\rm{exp}} \propto E_{\rm{exp}}^{1/2} M_{\rm{ej}}^{-1/2}$. Based on our prior light curve analysis, SN~2022ngb possesses a low-mass hydrogen envelope and, according to models from \cite{Nagy2016aap}, was produced by a relatively energetic explosion, making it comparable to SN~1993J, SN~2020acat, and SN~2024abfo \citep{Medler2022MNRAS}. These physical parameters collectively imply a moderate expansion velocity for the outer hydrogen layer, a prediction that is in agreement with our direct velocity measurements.

Neither hydrogen nor helium lines serve as reliable probes of the photosphere. Instead, \Feii lines, which originate deeper within the ejecta, can be used to trace the photospheric velocity \citep{Dessart2005AA}. Using this method, the photosphere velocity was approximately determined by using the \Feii $\lambda 5018$ line velocity at $-$2.3 days from \textit{V}-band maximum, giving a value of approximately 9000 $\mathrm{km \, s^{-1}}$. Here, we used it as a proxy for the photospheric velocity, which enabled us to perform a simple estimation of the explosion parameters, as discussed in Sect. \ref{LCM}.

\subsection{Nebular spectra}
The nebular phase spectroscopy data of CCSNe are typically used to estimate the parameters of their progenitor star and to constrain the explosion. At $+$95.4 days, the spectra of SN~2022ngb entered the nebular phase. This transition is marked by the emergence of prominent forbidden emission lines, such as \Oi $\mathrm{\lambda \lambda}$6300, 6364 and \Caii\ $\mathrm{\lambda \lambda}$7291, 7323, and the concurrent fading of the \Ha feature. From our observational data, we identified a split-peaked structure in both the \Oi $\mathrm{\lambda \lambda}$6300, 6364 lines, and this feature became more prominent throughout the spectral evolution. The red and blue wings, offset from the rest wavelength, correspond to a similar velocity of approximately 1000 $\mathrm{km \, s^{-1}}$. Also, the \Caii\ $\mathrm{\lambda \lambda}$7291, 7323 doublet develops an increasingly redshifted profile. These features indicate an aspherical ejecta distribution \citep{Milisavljevic2010ApJ, Valenti2011MNRAS}, similar to the asymmetric explosion features suggested by \cite{Fang2024NatAs}. These asymmetric features will be discussed in more detail in Sect. \ref{sec_asymm}.

\subsubsection{Oxygen mass and \Caii / \Oi ratio} \label{OCa}
The ejected mass of oxygen serves as a key indicator for the progenitor mass. As suggested by \cite{Uomoto1986ApJ}, a lower limit for the neutral oxygen mass can be estimated under the high-density limit ($N_e > 10^6\, \mathrm{cm^{-3}}$). This mass is calculated using the following relation,
\begin{equation}
M_O = 10^8 D^2 f(\Oi) \times \mathrm{exp}\left(\frac{2.28}{T_4}\right) \, \msun \, .
\end{equation}
where $M_O$ is the mass of neutral oxygen, $f(\mathrm{\Oi})$ is the integrated flux of the \Oi\ $\mathrm{\lambda \lambda}$6300, 6364 doublet in units of erg s$^{-1}$ cm$^{-2}$, $D$ is the distance in Mpc, and $T_4$ is the temperature of the oxygen-emitting region in units of $10^4$ K. Ideally, $T_4$ should be determined from the flux ratio of the auroral line \Oi\ $\mathrm{\lambda}$5577\AA\ to the nebular \Oi\ doublet. However, in our spectra, the \Oi\ $\mathrm{\lambda}$5577 line is too faint for a reliable measurement compared to the \Oi\ doublet. Furthermore, the \Oi\ $\mathrm{\lambda}$5577 line is potentially blended with other features, such as \Feii. Given these limitations, we adopted the approximation for the high-density, low-temperature regime by setting $T_4 = 0.4$ (corresponding to $T \approx 4000$ K), following the approach of \cite{Elmhamdi2004A&A}.

However, \cite{Medler2022MNRAS} noted that this method can be unreliable, primarily due to the temporal evolution of the oxygen emission flux. A more robust approach to establish a limit to oxygen mass, therefore, would be the application of this method to multiple epochs. Adopting the low-temperature approximation and using the measured \Oi\ $\mathrm{\lambda \lambda}$6300, 6364 fluxes of $2.18 \times 10^{-14}$ erg s$^{-1}$ cm$^{-2}$ at $+$162 days since explosion, we calculated an oxygen mass of $M_O \approx 0.68\ \msun$. Furthermore, we noted the persistent detection of the \Oinforb\ triplet near 7775 \AA\ in the nebular spectra of SN~2022ngb. The presence of this feature, which arises from the recombination of singly ionized oxygen \citep{Gangopadhyay2018mnras}, implies that a fraction of the oxygen remains ionized. Consequently, the estimate of $0.68\ \msun$ derived from neutral oxygen at +162 days should be regarded as a lower limit to the total ejected oxygen mass. For comparison, \cite{Elmhamdi2004A&A} showed that for typical SESNe, the ejected oxygen mass generally ranges from 0.2 $\msun$ to 1.4 $\msun$. The oxygen mass of SN~2022ngb thus falls within this distribution, and is comparable to those of SN~1993J ($\approx 0.5 \, \msun$; \citealt{Houck1996ApJ}) and SN~2015as ($\approx 0.45 \, \msun$; \citealt{Gangopadhyay2018mnras}), while being larger than that of SN~2011dh ($\approx 0.2 \, \msun$; \citealt{Sahu2013mnras}).

The oxygen observed in SN ejecta is predominantly synthesized during the hydrostatic nuclear burning stages within massive stars. Consequently, the total mass of ejected oxygen is directly linked to the initial main-sequence mass of the progenitor, as this parameter dictates the extent and efficiency of the nuclear burning phases. The nucleosynthesis models of \cite{Thielemann1996ApJ} provide a quantitative relationship between these two parameters. Their simulations predict that progenitors with initial main-sequence masses of 13, 15, 20, and 25 \msun\ will eject approximately 0.22, 0.43, 1.48, and 3.00 \msun\ of oxygen, respectively. By interpolating within the theoretical yields from \cite{Thielemann1996ApJ}, an oxygen mass of $\approx$0.68~\msun\ points to a progenitor with a main-sequence mass slightly exceeding 15 \msun. This inference can be cross-checked for consistency. The models of \cite{Thielemann1996ApJ} also indicate that a progenitor of $\sim$15 \msun\ evolves to form a helium core of approximately 4 \msun\ before collapse. This theoretical value matches our independent estimate of the helium core mass. Assuming a 1.5~\msun\ NS remnant and incorporating our earlier estimate of the total ejecta mass, we derived a helium core mass for SN~2022ngb in the range of 4.5 to 5.0 \msun, which is still consistent with the theoretically predicted He-core mass for a main-sequence progenitor of around 15 \msun. 

\cite{Fransson1989ApJ} suggested that the flux ratio of the \Caii\ $\mathrm{\lambda \lambda}$7291, 7323\AA\ doublet to the \Oi\ $\mathrm{\lambda \lambda}$6300, 6364\AA\ doublet provides a semi-quantitative method for estimating the main-sequence mass of the progenitor. The physical basis for this method is that oxygen is primarily formed during the hydrostatic evolution, whereas calcium is synthesized during the explosive nucleosynthesis. Therefore, the calcium yield is largely independent of the pre-SN evolution. Thus, a higher \Caii / \Oi ratio points to a less massive progenitor. \cite{Elmhamdi2004A&A} found that the \Caii / \Oi ratio becomes stable at late epochs (typically at $+$150 days past explosion), remaining nearly constant for an extended period. In the case of SN~2022ngb, the \Caii / \Oi ratio at $+$162 and $+$252 days post-explosion is indeed observed to be nearly constant, with a value of approximately 0.49. This value is comparable to those of SN~2020acat ($\approx 0.5$) and SN~1993J ($\approx 0.5$), suggesting that SN~2022ngb had a progenitor with a mass in the range from 14 to 18 \msun. This is in agreement with our previous estimates. However, it should be noted that the underlying assumptions for this method are not always robust. Furthermore, the \Caii\ $\mathrm{\lambda \lambda}$7291, 7323\AA\ feature is often blended with nearby \Oii\ and \Feii\ lines, which introduces significant uncertainty into the flux measurement. In such cases, the \Caii / \Oi ratio may not be a reliable probe of the progenitor mass. This disagreement arises in the estimation of SN~2008ax, which shows a high \Caii / \Oi ratio despite its massive progenitor \citep{Taubenberger2011MNRAS}. 

\subsubsection{Modelling spectra in the nebular phase} 
\label{sec_subsubnebmdl}
\begin{figure}[htbp]
\centering
\includegraphics[width=1.0\linewidth]{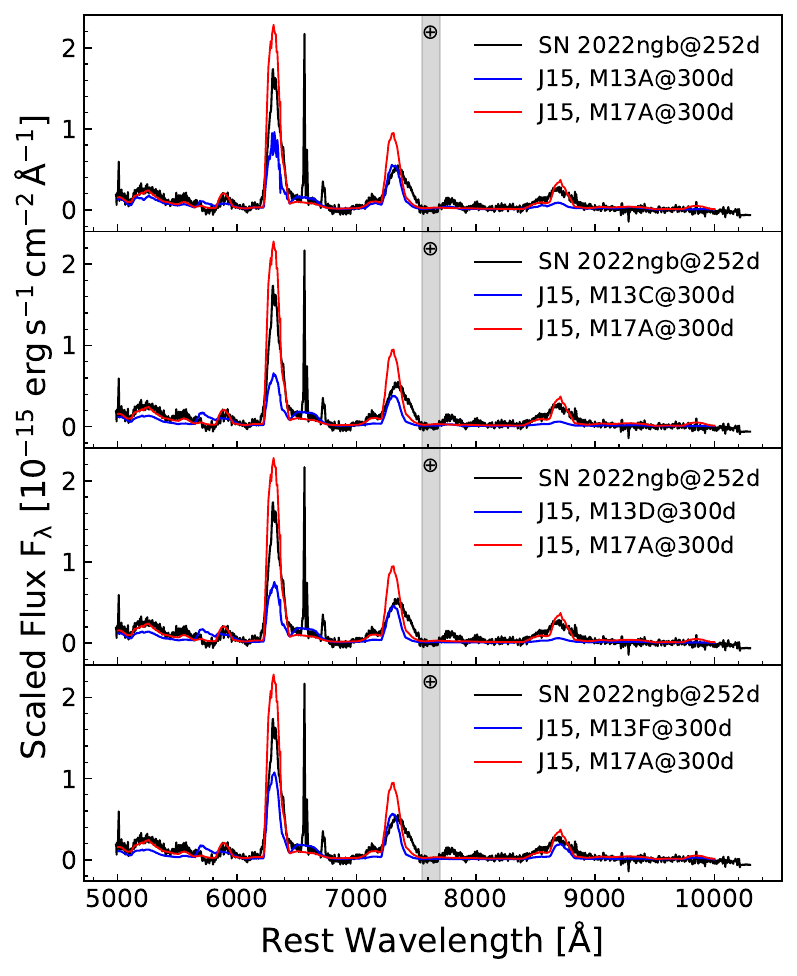}
\caption{Nebular phase spectrum of SN~2022ngb at +252 days post-explosion (black line), compared with synthetic spectra taken from the SN IIb models of \cite{Jerkstrand2015aap}. The blue lines represent various models for a $M_\mathrm{{ZAMS}}=$13 $\mathrm{M_\odot}$  progenitor, while the red line shows a model for a 17 $\mathrm{M_\odot}$ progenitor. The narrow lines in the spectra are the emission lines from the \Hii region of the host galaxy.
}
\label{spec_cmp_mdl_neb}
\end{figure}

We compared a nebular-phase spectrum of SN~2022ngb, obtained at $+$252 days past explosion ($+$227 days after the \textit{V}-band maximum), by comparing it with synthetic spectra generated using the \texttt{SUMO} code \citep{Jerkstrand2015aap, Barmentloo2024MNRAS}. The comparison is presented in Fig.~\ref{spec_cmp_mdl_neb}. The narrow emission lines in the observed spectrum, attributed to \Hii\ regions in the host galaxy, were excluded from this analysis. The models employed are based on the stellar evolution calculations of \citet[hereafter WH07]{Woosley2007PhR} for progenitors with $M_\mathrm{{ZAMS}}$ of $13 \, \msun$ and $17 \, \msun$. These progenitors correspond to ejecta masses ranging from $2.1 \, \msun$ to $3.5 \, \msun$, a range that encompasses our estimate for SN~2022ngb in Sect. \ref{OCa}. The models assume an initial $^{56}$Ni mass of $0.075 \, \msun$. To facilitate a direct comparison, synthetic spectra were scaled to the distance of SN~2022ngb and adjusted for radioactive power at the observed epoch. 

To constrain the progenitor mass of SN~2022ngb, we compared its nebular spectrum with a suite of models for $M_\mathrm{{ZAMS}} = 13 \, \msun$ and $M_\mathrm{{ZAMS}} = 17 \, \msun$ stars from \cite{Jerkstrand2015aap}, namely \texttt{M13A}, \texttt{M13C}, \texttt{M13D}, \texttt{M13F}, and \texttt{M17A}. Our analysis proceeded by first eliminating models with incompatible physical assumptions. A comparison between \texttt{M13C} (with dust) and \texttt{M13D} (without dust) reveals that dust has a negligible effect on the optical spectrum, and it is thus excluded from further consideration. Similarly, comparisons with \texttt{M13D} (with molecular cooling) and \texttt{M13F} (without molecular cooling) indicate that the latter provides a superior fit. This leaves \texttt{M13F} as the most representative model among the lower-mass options. While this model provides a reasonable overall fit, it underpredicts the flux of the \Oi\ and overestimates the strength of the \Nii\ emission lines. This discrepancy suggests that the true progenitor mass is higher, above $M_\mathrm{{ZAMS}} = 13 \, \msun$. Conversely, the \texttt{M17A} model ($M_\mathrm{{ZAMS}} = 17 \, \msun$) significantly overpredicts the strength of the \Oi\ emission, allowing us to set a firm upper limit on the progenitor mass. By interpolating between these lower and upper bounds, we estimate the progenitor mass for SN~2022ngb to be in the range of $M_\mathrm{{ZAMS}} \approx 15-16 \, \msun$, a result that is consistent with our previous discussion.
\begin{figure*}[htbp]
\centering
\includegraphics[width=1.0\linewidth]{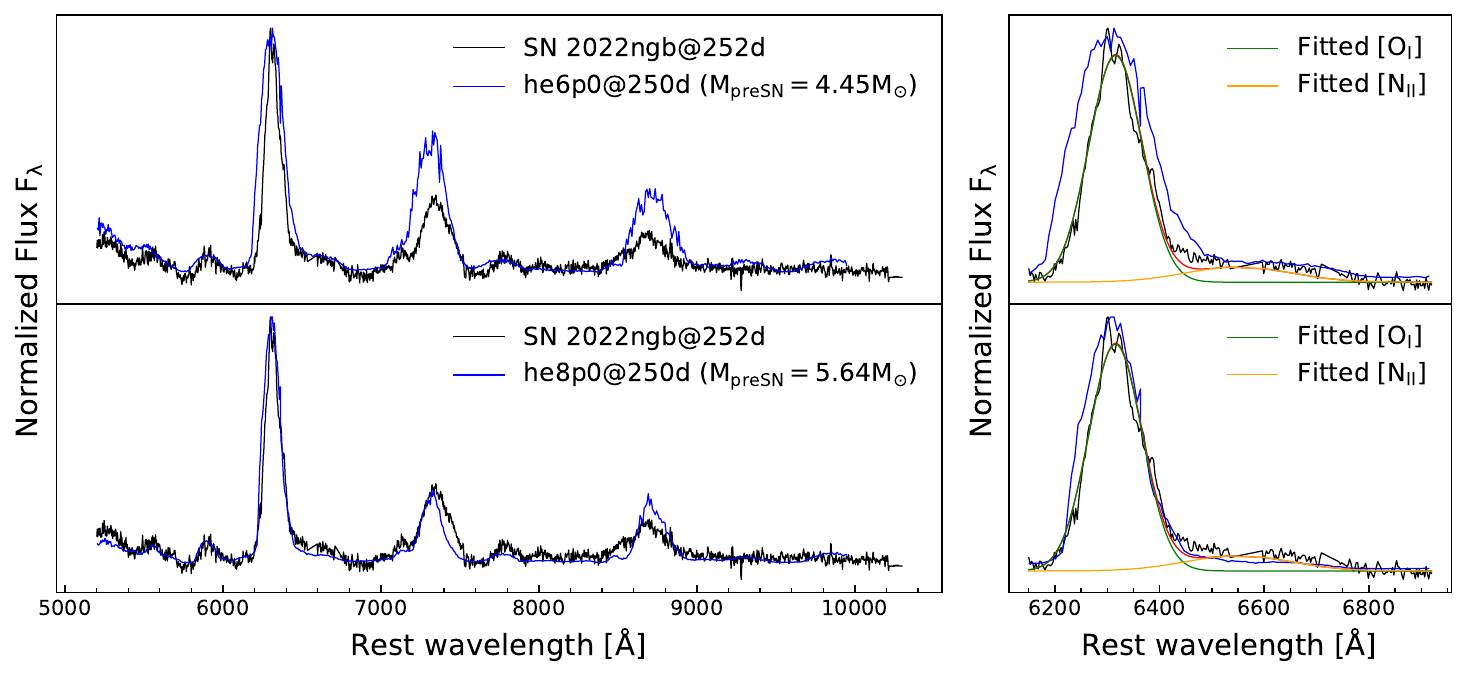}
\caption{Nebular phase spectrum of SN~2022ngb at +252 days from explosion (black line), compared with synthetic spectra from the He-star models of \cite{Barmentloo2024MNRAS}. All spectra are normalized to their maximum flux and have the continuum removed. Left panel: Comparison of the nebular spectrum of SN~2022ngb with He-star models of varying $M_\mathrm{{preSN}}$. Right panel: A detailed comparison of the \Oi and \Nii line profiles. The green and orange solid lines represent Gaussian fits to the \Oi and \Nii components, respectively. The emission lines from the \Hii region of the host galaxy is removed for a better comparison.}
\label{spec_neb_nii}
\end{figure*}

As mentioned in Sect. \ref{speccmp_sec}, the nebular \Nii lines can serve as a probe to constrain the mass of the progenitor. The origin of the nitrogen lies in the CNO cycle, where the $^{14}$N($p$, $\gamma$)$^{15}$O reaction is the rate-limiting step, leading to the accumulation of $^{14}$N. We assume a scenario in which the hydrogen envelope of the progenitor was partially stripped, leaving nitrogen-rich material from the H-burning shell mixed at the surface of the dense, He-rich core. During the subsequent helium-shell burning, this $^{14}$N is efficiently converted to $^{18}$O via the $^{14}$N($\alpha$, $\gamma$)$^{18}$F reaction, which subsequently $\beta$-decays to $^{18}$O. A more massive He-core implies higher temperatures and thus a faster reaction rate. To perform a more precise flux estimation of \Nii\ and avoid the blending with other lines (e.g., \Feii and \Ha lines; \citealt{Barmentloo2024MNRAS}), we use a late-time spectrum of SN~2022ngb taken at $+$252 days post-explosion. At this advanced epoch, the contaminating effects of \Feii and \Ha are negligible.

The mass loss from the helium core of the progenitor is another crucial factor. For Type Ib/Ic SNe, which arise from fully stripped He stars, strong stellar winds significantly reduce the final $M_\mathrm{{preSN}}$. In contrast, for SNe IIb such as SN~2022ngb, the presence of a residual hydrogen envelope implies that mass loss from the underlying He core was less significant. This view is supported by the work of \cite{Dessart2023AA}, who suggested that models with the same $M_\mathrm{{preSN}}$ produce similar nebular spectra, regardless of their initial mass. Therefore, we adopt the simplifying assumption that the He core of the progenitor experienced a negligible mass loss ($\frac{\mathrm{d}M}{\mathrm{dt}} = 0$, as suggested by \citealt{Dessart2023AA}), making its pre-SN mass equivalent to its He-core mass. Following \cite{Barmentloo2024MNRAS}, we compared our spectrum with their \texttt{he6p0} and \texttt{he8p0} models, which have $M_\mathrm{{preSN}} = 4.45~\msun$ and $5.64~\msun$, respectively. This mass range is consistent with our previous estimates. The comparison is presented in Fig.~\ref{spec_neb_nii}.

Overall, both the \texttt{he6p0} and \texttt{he8p0} models from \cite{Barmentloo2024MNRAS} provide a good fit to our observed spectrum. Focusing specifically on the \Nii lines, our data is bracketed by the two models: the \texttt{he6p0} model predicts a stronger \Nii emission than observed, whereas the emission in the \texttt{he8p0} model is slightly weaker. Furthermore, in the more massive \texttt{he8p0} model, the \Nii feature at $+$252 days is still significantly blended with \Feii emission. This blending implies that the true \Nii flux of the models is even lower than what is indicated by the blended feature, reinforcing the conclusion that this model underpredicts the nitrogen abundance in SN~2022ngb.

To obtain a more accurate estimate of the $M_\mathrm{{preSN}}$, we utilized the flux ratio of \Nii to the total flux in the 5000--8000 \AA\ wavelength range. This ratio serves as a sensitive probe for $M_\mathrm{{preSN}}$ and is defined as \citep{Barmentloo2024MNRAS}
\begin{equation}
f_{\mathrm{\Nii}} = \frac{F_{\mathrm{\ion{N}{ii},\,fit}}}{\int_{5000 \AA}^{8000 \AA} F_{\lambda} d \lambda} \times 100 \, .
\end{equation}
To prevent the \Feii and \Ha lines from skewing the estimate, we used data from the late nebular phase, during which the contribution from these lines is expected to be negligible \citep{Jerkstrand2015aap, Barmentloo2024MNRAS}. 

For SN~2022ngb, this ratio is $\sim$2.1 at +252 days past explosion. We compared this value to the \texttt{he8p0} and \texttt{he6p0} models from \cite{Barmentloo2024MNRAS}, which yield $f_{\mathrm{\Nii}}$ values of approximately 1.5 and 2.5, respectively. By interpolating between these models based on our measured $f_{\mathrm{\Nii}}$ value, and incorporating the mass of the hydrogen-rich shell derived from other models \citep[e.g.][]{Piro2021ApJ, Nagy2016aap}, we estimated the total progenitor mass to be $4.7$ \msun, consistent with the outcomes of the light-curve fits, which indicate an ejecta mass of $\sim 3$ \msun. Summing the ejecta mass with a typical NS remnant mass of 1.6$-$1.8 \msun\ yields a comparable total mass. Furthermore, the explosion energies of these models are in good agreement with that estimated for SN~2022ngb.

\section{Discussion}
\label{sec_phys}

\subsection{The early shock-cooling emission}
\label{sec_shockcooling}
\begin{figure}[htbp]
\centering
\includegraphics[width=1.0\linewidth]{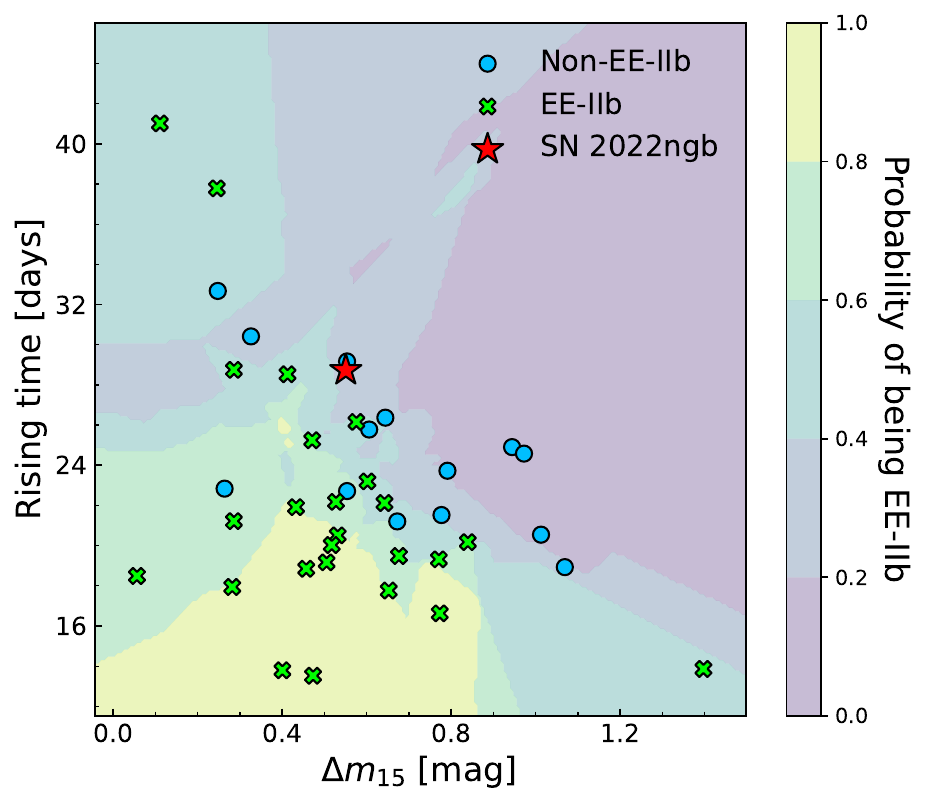}
\caption{Magnitude decline rate vs. rise time in SNe IIb in the ATLAS \textit{o} bands. The cyan points represent SNe IIb with no prominent early emission (Non-EE-IIb), and the green points stand for those with prominent early emission (EE-IIb). The red star indicates the position of SN~2022ngb in the diagram. The different colored regions in the plots represent the K-NN-derived probability that an event belongs to the EE-IIb class.}
\label{dm15vsrisingtime}
\end{figure}

Early-time observations of SNe IIb provide crucial insights into the properties of the hydrogen-rich envelope of the progenitor star. The early emission (EE) is understood to arise from the interaction of the shock with this extended envelope, manifesting as a rapidly rising light curve followed by a declining cooling phase.
For SN~2022ngb, the light curve feature in this phase is similar to SN~2008ax \citep{Pastorello2008mnras}, SN~2015as \citep{Gangopadhyay2018mnras}, and SN~2020acat \citep{Medler2022MNRAS}, which present a faint shock cooling feature, suggesting a low-mass, hydrogen-rich envelope with small radii.
Acquiring observational coverage of this brief period is notoriously difficult. \cite{Ayala2025A&A...701A.128A} studied the light curve properties of SNe IIb with and without EE based on ATLAS archival data. Their analysis indicates no statistically significant difference between the rise-time and post-peak decline rate distributions between the two populations.
Here, we utilized the rise times and decline rates from Table \ref{tab_appapplc} for a comparison with a sample of other SNe IIb events (both with and without EE). The outcomes of this comparison are presented in Fig.~\ref{dm15vsrisingtime}.

We found that SN~2022ngb overlaps with both EE and non-EE SNe IIb in terms of rise time and decline rate ($\Delta m_{15}$). However, given the lack of statistically significant separation between these two populations \citep{Ayala2025A&A...701A.128A}, making accurate classification based on these parameters alone is difficult. To quantify the classification probability within this overlapping framework, we employed the K-nearest neighbors (K-NN)\footnote{\url{https://scikit-learn.org/stable/modules/generated/sklearn.neighbors.KNeighborsClassifier.html}} approach. Given the limited sample size (15 non-EE-IIb and 25 EE-IIb events), we adopted $k=5$ to balance local sensitivity with sample robustness. The analysis yields a probability of 60\% for non-EE-IIb and 40\% for EE-IIb. This result suggests that while SN~2022ngb is more likely a non-EE-IIb event, it could represent a transitional event between the two subclasses. This is consistent with the theoretical framework of \cite{Chevalier2010ApJL}, which suggests that a faint or absent shock-cooling phase is indicative of a thin, low-mass envelope. Therefore, these features imply that SN~2022ngb is likely a cIIb, originating from a progenitor with a less extended envelope, analogous to SN~2022crv.

To further analyze the luminosity evolution, we followed \cite{Rabinak2011apj}, where the luminosity is expressed as
\begin{equation}
    L(t) \propto \left(\frac{E}{E_0}\right)^{\gamma_1} \left(\frac{R}{R_0}\right) f_{\rho}^{-\gamma_2} \left( \frac{M}{M_{\odot}} \right)^{-\gamma_3} \left(\frac{\kappa}{\kappa_0}\right)^{-\gamma_4} \left(\frac{t}{t_0}\right)^{-\gamma_5} \, .
\end{equation}
Here, $\gamma_1$, $\gamma_2$, $\gamma_3$, $\gamma_4$, and $\gamma_5$ are positive constants that depend on the density profile. In this case, the decline rate $\dot L(t) \propto - R M^{-\gamma_3} t^{-\gamma_5 - 1}$. The best-fit result present in Table \ref{apptab_snlcfittingparams} of SN~2022ngb shows a steep decline rate during the first ~10 days, suggesting a low-mass envelope. In comparison to SN~2015as \citep{Gangopadhyay2018mnras}, which had a similar envelope radius, its shell luminosity declined to $10^{41.8}$ erg s$^{-1}$ at about 9 days after explosion, showing a much slower early decline rate than SN~2022ngb. The luminosity of the shell component in the \texttt{LC2} model of SN~2022ngb dropped to be negligible at nearly the same epoch, which supports the conclusion of an extremely low-mass envelope. This comparison between SN~2022ngb and SN~2015as would indicate that the envelope of SN~2022ngb would be more compact and less massive than SN~2015as.

\begin{figure}[htbp]
\centering
\includegraphics[width=1.0\linewidth]{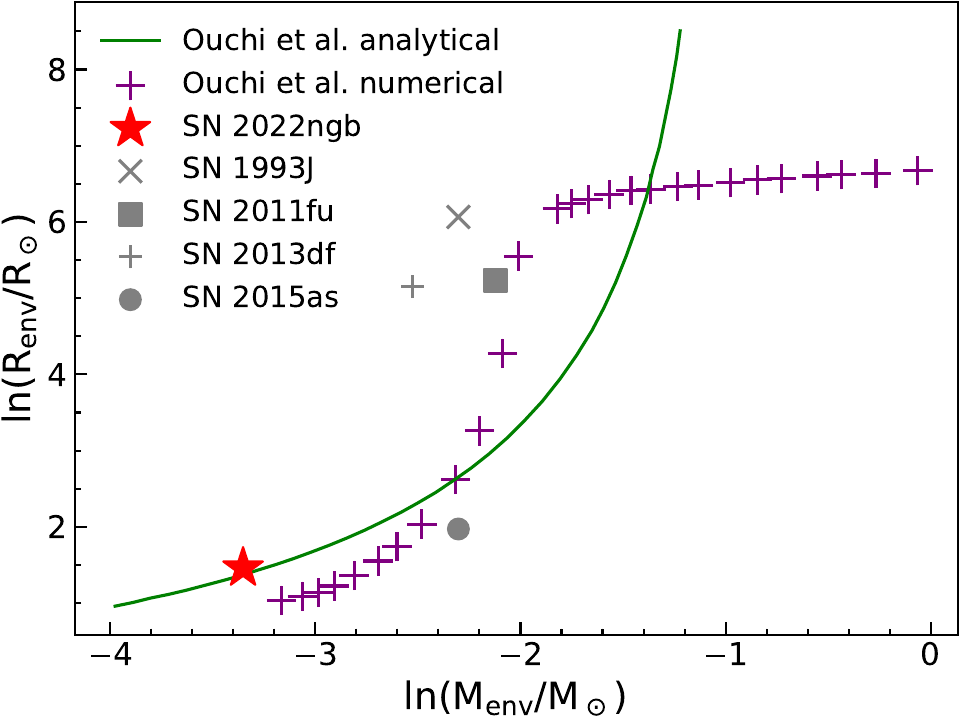}
\caption{$M_\mathrm{env}$ and $R_{\mathrm{env}}$ relations of SN~2022ngb and other typical SNe IIb compared to the \citet[OM17]{Ouchi2017apj} model. The green solid line stands for the analytical solution of OM17 model, while the purple cross point stands for their numerical solution. The red star point is the position of SN~2022ngb, the other gray markers stands for other SNe IIb listed in the legend.}
\label{OM17_cmp}
\end{figure}

We also checked the plausibility of the envelope properties by examining the internal physics of the progenitor. The properties of the envelope are determined by the structure of the progenitor and the mass loss history. Through an analytical model and numerical simulations by \cite{Ouchi2017apj}, they described the relationship between $M_{\mathrm{env}}$ and $R_{\mathrm{env}}$ for progenitors of SNe IIb. Considering SN~2022ngb and other typical SNe IIb we used, we plot the corresponding $M_{\mathrm{env}}$-$R_{\mathrm{env}}$ values for each target in Fig.~\ref{OM17_cmp}. In this figure, SN~2022ngb and other comparisons fits the \cite{Ouchi2017apj} model well. This indicates that the results of envelope properties of SN~2022ngb is reasonable.

\subsection{Photospheric evolution of \Ha}
\label{sec_haevo}

\begin{figure*}[htbp]
\centering
\includegraphics[width=1.0\linewidth]{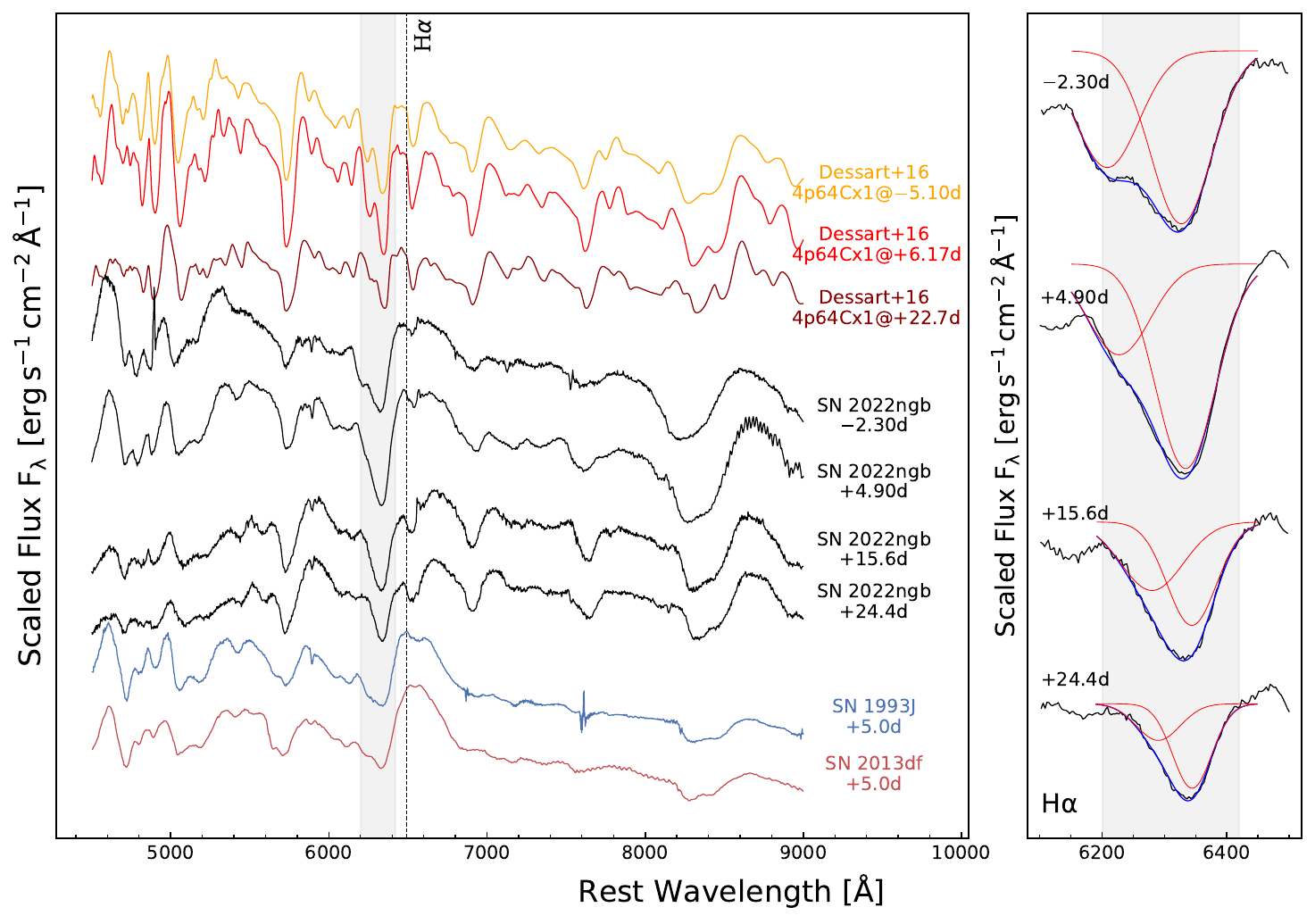}
\caption{Left panel: Comparison of spectra of SN~2022ngb obtained near maximum light with spectra of the SNe IIb 1993J and 2013df, and synthetic spectra from the NLTE models of \cite{Dessart2016MNRAS}. The models shown correspond to a progenitor with a pre-explosion mass of 4.64 \msun. The rest wavelength position of \Ha is marked with a dotted vertical line. Right panel: Details of the evolution of the H$\alpha$ absorption. Its profile is fitted with a double-Gaussian function (blue solid line), whose individual components are shown with red solid lines.
}
\label{spec_hacmp}
\end{figure*}

The detailed evolution of the \Ha profile is shown in Fig.~\ref{spec_hacmp}, compared to two representative SNe IIb with similar profiles and models for a progenitor with pre-explosion mass of 4.64 \msun from \cite{Dessart2016MNRAS}. The spectrum of SN~2022ngb at $-$2.3 days from \textit{V}-band maximum shows a two-component blue-shifted absorption for \Ha. Using a Gaussian fit, we determined that the central wavelength of the blue wing component is 6220 \AA, and that of the red wing component is 6320 \AA. The red component is the absorption of the \Ha~P~Cygni profile, while the origin of the blue component is unclear. In our initial fit, we reproduced the broad \Ha feature as a combination of \Ha and \Siliconii. We observe that this broad feature becomes narrower as the blackbody temperature drops from approximately 6500~K in the earlier phase to 5500~K.

To improve our understanding of the blue wing absorption, we compare the spectra with the NLTE radiative transfer models from \cite{Dessart2016MNRAS}. The models exhibit characteristics that are in fair agreement with our observations at similar epochs. The primary discrepancy, observed in the \Hei features, can be attributed to a higher ejecta density and opacity, as previously discussed. We note that in the \texttt{4p64Cx1} model, the \Siliconii lines are prominent throughout the spectral evolution, and the velocity inferred from their absorption minima shows a decelerating trend. Furthermore, the absorption component of the P~Cygni profile becomes progressively narrower, a behavior that is also consistent with our observational data.

We noted a mismatch at later phases ( $+$6.17 days and $+$22.7 days from \textit{V}-band maximum in the \texttt{4p64Cx1} model), where the \Siliconii absorption remains prominent in the model but has nearly vanished in our data. A similar behavior was observed in SN 2015as \citep{Gangopadhyay2018mnras}. We suggest this discrepancy arises from the blending of the decelerating \Siliconii line with the increasingly strong \Ha profile, which effectively masks the \Siliconii feature. To isolate this underlying blue wing component, we performed a double-Gaussian fitting to the observed profile. This analysis confirmed the persistence of a blue-shifted absorption component, revealing a flux comparable to that predicted by the \texttt{4p64Cx1} model. Furthermore, the fitting process enabled us to measure the velocity of the \Siliconii line that decreased from 13500 $\mathrm{km \, s^{-1}}$ at $-$21.1 days to 6300 $\mathrm{km \, s^{-1}}$ at $-$2.3 days, and further to 6000 $\mathrm{km \, s^{-1}}$ at $+$4.9 days. This trend is consistent with the typical velocity evolution of inner ejecta layers being revealed as the photosphere recedes. In addition to line blending, other physical effects contribute to the weakening of the \Siliconii absorption. The temperature drop from 5500~K to 4500 K would further reduce the \Siliconii ionization fraction, along with the decreasing density of the expanding ejecta and the recession of the photosphere into deeper, slower moving layers.

We noted that this feature has also been observed in other SNe IIb, such as SN~1993J \citep{Barbon1995aaps} and SN~2013df \citep{Milisavljevic2013apj}, as evidence of either an aspherical hydrogen distribution \citep{Barbon1995aaps} or a distinct, high-velocity hydrogen layer \citep{Branch2002ApJ, Milisavljevic2013apj}. The evolving absorption profile around 6200 \AA\ in SN 2022ngb closely resembles that seen in SN~2015as \citep{Gangopadhyay2018mnras}, suggesting a shared physical mechanism. Our analysis, consistent with the findings for SN~2015as, indicates that the double-trough profile results from a blend of the P~Cygni absorption from both \Siliconii and \Ha. This interpretation is compelling because a separate high-velocity hydrogen component might be expected to show a velocity evolution parallel to the primary \Ha line, as observed in SN 2020cpg \citep{Medler2021MNRAS}, a behavior which has not been observed in our data. Nevertheless, alternative explanations, such as an aspherical hydrogen distribution (e.g., in a toroidal geometry), cannot be entirely ruled out. 

\subsection{Evidence for asymmetric ejecta from nebular spectroscopy}
\label{sec_asymm}

\begin{figure}[htbp]
\centering
\includegraphics[width=1.0\linewidth]{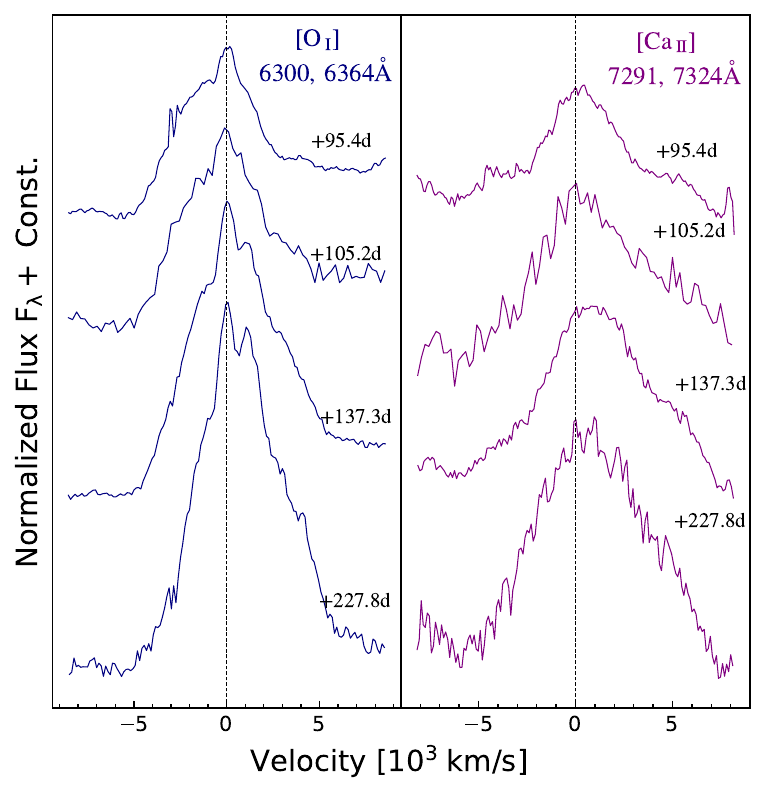}
\caption{Evolution of the profiles of the \Oi $\lambda\lambda$6300, 6364 \AA \, doublet (left panel), and the \Caii $\lambda\lambda$7291, 7324 \AA \, doublet (right panel). The rest wavelength for each line is marked by a vertical dashed line. The observational epoch is indicated to the right of each spectrum.}
\label{spec_evo_neb}
\end{figure}

The evolution of the prominent lines during the nebular phase, in particular \Oi $\mathrm{\lambda \lambda}$6300, 6364 \AA, and \Caii $\mathrm{\lambda \lambda}$7291, 7323 \AA~ (Fig. \ref{spec_evo_neb}), reveals that SN~2022ngb behaves similarly to other well-studied Type IIb events, such as SN~2015as \citep{Gangopadhyay2018mnras} and SN~2020acat \citep{Medler2022MNRAS}. 
The nebular spectra of SN~2022ngb are dominated by the forbidden emission doublet \Oi\ $\mathrm{\lambda \lambda}$6300, 6364 \AA. As the ejecta expand and become optically thin, these lines develop a prominent, asymmetric double-peaked profile. The profile is characterized by a strong component centered close to the rest wavelength at 6300 \AA\ (as indicated by the dashed line in the figure), a distinct secondary peak redshifted to a velocity of approximately $+$1000 $\mathrm{km \, s^{-1}}$, and a less prominent, blueshifted peak at a velocity of roughly $-$1000 $\mathrm{km \, s^{-1}}$. This line profile is reminiscent of those observed in other Type IIb events, such as SN~2008ax \citep{Pastorello2008mnras} and SN~2015as. Several physical mechanisms can produce such a profile. One possibility is an intrinsic asymmetry in the ejecta, which could arise from large-scale convective instabilities in the oxygen layer \citep{Kifonidis2006A&A}, an asymmetric explosion creating an inhomogeneous oxygen distribution via Rayleigh-Taylor instability \citep{Baal2024MNRAS} or a fundamentally aspherical (e.g., equatorial) distribution of oxygen \citep{Taubenberger2009MNRAS, Fang2024NatAs}. 

If we consider the asymmetric explosion model presented by \cite{Fang2024NatAs}, the profile of SN~2022ngb is best interpreted as originating from an aspherical oxygen distribution, possibly a toroidal structure around the equator. The observed velocity split of $\sim$1000 $\mathrm{km \, s^{-1}}$ is moderate, suggesting that our line of sight is oriented at an intermediate angle relative to the axis of symmetry, rather than being fully equatorial. A purely equatorial view would likely produce a larger velocity split, while an extremely low viewing angle might not show the evolution feature of the red wing of \Oi\ $\mathrm{\lambda \lambda}$6300, 6364 \AA. This interpretation is further supported by the temporal evolution of the profile: as the spectrum transitions into the nebular phase, the increasing prominence of the red-wing peak relative to the blue-wing component indicates that the ejecta are becoming progressively more transparent, allowing for more emission from the receding side to be observed.

The \Caii\ $\mathrm{\lambda \lambda}$7291, 7323 \AA\ lines are a dominant feature in the nebular spectra of SN~2022ngb (Fig. \ref{spec_evo_neb}). The dashed line represents the rest wavelength (7291 \AA) of the primary component of the doublet. This feature shows a broad, rounded peak, somewhat different from the sharp peaks observed in SN~2008ax and SN~2011dh, but similar to those of SN~2015as and SN~2020acat. This suggests a large velocity dispersion of the emitting material. \citet{Medler2022MNRAS} proposed that an energetic explosion could cause such a broad velocity distribution observed in SN~2020acat. Our analysis of the bolometric light curve suggests a relatively high explosion energy of approximately $1.3\times10^{51}\,\mathrm{erg}$, consistent with the higher energy case of SN~2020acat. Also, based on the simulations in \cite{Baal2024MNRAS}, which show that \Caii dynamics can be a good probe of the explosion geometry, the broad \Caii line profile is more likely due to an asymmetric, powerful explosion. This conclusion is consistent with our results from the previous discussion. Furthermore, we observe that the peak of the \Caii\ $\mathrm{\lambda \lambda}$7291, 7323 \AA\ feature is gradually redshifted over time. This can be explained by the same mechanism responsible for the increasing visibility of the red wing of the \Oi\ $\mathrm{\lambda \lambda}$6300, 6364 \AA\ doublet. 

Furthermore, the \Caii\ $\mathrm{\lambda \lambda}$7291,7323 \AA\ doublet, which traces the bulk ejecta and overall explosion dynamics, exhibits a redshift of $\sim$1000 $\mathrm{km \, s^{-1}}$. This indicates a large-scale asymmetry where more material, including radioactive $^{56}$Ni, was propelled away from the observer. This anisotropic distribution of $^{56}$Ni can explain both the lower peak bolometric luminosity and the modest estimated $^{56}$Ni mass. 
Therefore, within the bipolar explosion framework of \cite{Fang2024NatAs}, the nebular profiles of SN 2022ngb indicate an intermediate viewing angle. This is supported by its \Oi velocity split, which places it between the nearly on-edge systems (e.g., SN 2011dh, SN 2015as) and the on-pole system, suggesting a potentially intermediate viewing angle. 

\subsection{Constraints on the progenitor and the explosion}
\label{sec_prog}

We use several approaches to determine the mass of the progenitor in Sect. \ref{sec_subsubnebmdl}. Taking these results and considering the research by \cite{Barmentloo2024MNRAS}, higher-mass helium stars generally follow the trend of having smaller radii. For SN~2022ngb, the combination of faint shock-cooling features in the bolometric light curve and a relatively high $M_\mathrm{preSN}$ value points to a progenitor with only a small and compact hydrogen-rich envelope. In this regard, SN~2022ngb is similar to SN~2022crv ($R \sim 3 \, R_{\odot}$, $M_\mathrm{{preSN}} = 5.3^{+0.3}_{-0.6}\, \msun\ $; \citealt{Gangopadhyay2023ApJ}) and SN~2003bg ($R \sim 1 \, R_{\odot}$, $M_\mathrm{{preSN}} = 4.0^{+0.4}_{-0.4}\, \msun\ $; \citealt{Soderberg2006ApJ}). These events are thought to originate from progenitors resembling relatively compact WR-like stripped stars, where the envelope has been almost entirely stripped, leaving only a thin residual shell behind. \cite{Yoon2017ApJ} suggested that progenitors lacking an extended hydrogen-rich envelope likely originate from binary systems with tight binding, which implies a short initial orbital period. This evidence indicates that the progenitor of SN~2022ngb was likely in a tightly bound binary system accordingly. 

In brief, the progenitor of SN~2022ngb was most likely a partially stripped star with a relative large $M_\mathrm{{preSN}} \approx 4.7$ \msun, corresponding to a $M_\mathrm{{ZAMS}}$ of approximately 15 \msun. Our model points to a relatively compact helium-rich core radius of about 1.44 $R_{\odot}$, surrounded by a thin, low-mass hydrogen-rich envelope extending to approximately 4.3 $R_{\odot}$. The properties of the progenitor is similar to SN~2008ax \citep{Crockett2008MNRAS} and likely to be stripped in a binary system. 

\section{Conclusion}
\label{sec_con}
In this work, we present comprehensive optical and NIR photometric \textit{BgcVroizJHK} and spectroscopic observations of the SN IIb~2022ngb. Our photometric data coverage began close to the explosion epoch, capturing a faint shock-cooling phase $\sim 3$ days from explosion. The photometric and spectroscopic data cover a period of over 270 and 250 days, respectively. A comparison of the \textit{V}-band absolute light curve with those of other well-studied SNe IIb reveals its luminosity is significantly lower than that of SN~1993J and SN~2011fu, whereas it is comparable to that of SN~2015as and SN~2024abfo. The color evolution of SN~2022ngb resembles those of SN~2008ax and SN~2015as, which are classified as cIIb SNe, a subclass characterized by a weak or absent initial shock-cooling peak and a smaller progenitor radius. The color indices indicate that SN~2022ngb has a cooler photospheric temperature evolution compared to typical SNe IIb. 

The peak bolometric luminosity of SN~2022ngb (derived by applying a blackbody correction) is $L_{\mathrm{Bol}} = 7.76^{+1.15}_{-1.00} \times 10^{41} \, \mathrm{erg \, s^{-1}}$. This value indicates a lower bolometric luminosity compared to most SNe IIb, placing it among low-luminosity events such as SN~1996cb and SN~2011ei. The rise time to peak bolometric luminosity is approximately 28.5 days, which is slightly longer than that of typical SNe IIb. A modeling of the light curve points to an ejecta mass in the range from 2.8 to 3.3 \msun\ and a relatively high explosion energy of around $1.4\times10^{51}\,\mathrm{erg}$. The progenitor of SN~2022ngb had a low-mass and compact hydrogen-rich envelope, with a mass from 0.03 to 0.08 \msun\ and a radius of less than 4 $R_{\odot}$. These properties point to a compact, partially stripped-envelope progenitor, possibly a lower-mass analogue to a WR-like stripped star. The synthesized $^{56}$Ni mass for SN~2022ngb is in the range 0.043-0.045 \msun, based on a modified Arnett-like model \citep{Arnett1982apj, Chatzopoulos2012apj} and a two-component model \citep{Nagy2016aap}. This $^{56}$Ni mass is comparable to that of SN~2024abfo, which exhibits a similar light curve.

The nebular spectra of SN~2022ngb are similar to those of SNe~2020acat and 2015as, both of which exhibit strong \Oi emission and relatively weak \Caii emission. The asymmetric profile of the \Oi lines is comparable to that seen in SNe~2015as and 2008ax, suggesting a globally spherical ejecta geometry with a toroidal or disk-like structure.
The red-shifted \Caii lines point to an asymmetric explosion. This asymmetry could imply an anisotropic distribution of radioactive material and could contribute to the observed lower luminosity if viewed from a direction with a deficit of $^{56}$Ni. Comparisons with synthetic spectra indicate that the progenitor of SN~2022ngb had a $M_\mathrm{{preSN}}$ of approximately $4.7$ \msun\ and a $M_\mathrm{{ZAMS}}$ of 15 to 16 \msun. Finally, a comparison with other Type-eIIb and cIIb SNe suggests that the progenitor was compact, similar to what was found in the cases of SNe~2008ax, 2022crv, and 2003bg. This implies an origin in a binary system, where interaction with a companion stripped most of its envelope, leaving a progenitor that resembles a compact Helium star.

Modern, large-scale, high-resolution survey projects are underway to discover a greater number of SNe. These efforts will enhance our understanding of their explosion mechanisms and the diversity of their progenitor systems. In the coming years, thousands of transients will be discovered by next-generation time-domain surveys such as the Vera C. Rubin Observatory's Legacy Survey of Space and Time \citep[LSST;][]{Ivezic2019ApJ}, the China Space Station Telescope \citep[CSST;][]{Gong2019ApJ}, and the Multi-channel Photometric Survey Telescope \citep[Mephisto;][]{Yuan2020SPIE11445E}. The LSST, in particular, is expected to detect up to 10 million SNe over its 10-year mission, vastly increasing the known sample. Meanwhile, the Mephisto survey, consisting of two 50-cm telescopes and a 1.6-m telescope, will enable simultaneous observations in both the \textit{ugi} and \textit{vrz} bands. This capability is crucial for accurately retrieving the true SED and color information of SNe. Collectively, these facilities will provide us with more precise and crucial insights into the mechanisms of SNe and the evolution of their progenitors.

\section*{Data availability}
Photometric data for the Type IIb SN~2022ngb presented in this study are available at the CDS via \url{https://cdsarc.cds.unistra.fr/viz-bin/cat/J/A+A/706/A271}. Our observations are available via the Weizmann Interactive Supernova Data Repository (WISeREP; \citealt{Yaron2012pasp}).

\bibliographystyle{aa}
\bibliography{refs.bib}

\begin{appendix}
\onecolumn

\section{Photometric tables}
\label{app_compl}
\begin{table*}[htbp]
    \centering
    \caption{Basic information of SN\,2022ngb compared with other well-observed Type~IIb SNe.}
    \resizebox{\columnwidth}{!}{
    \begin{tabular}{cccccccccc}
        \hline
        SN & Explosion Epoch & Redshift & Distance & $E(B-V)_\mathrm{MW}$ & $E(B-V)_\mathrm{Host}$ & $^{56}$Ni Mass& $M_{\mathrm{ej}}$ & $M_{\mathrm{preSN}}^* $&Source \\
         & [MJD] & $z$& [Mpc] & [mag] & [mag] & [\msun] & [\msun] & [\msun] & \\
        \hline
        1993J & $49072.0\pm0.5$ & $-$0.00113 & $3.6\pm0.3$ & 0.069 & 0.11 & 0.1& 1.9& 3.1&1,2,3 \\
        2008ax & $54528.8\pm0.2$ & 0.00456 & $9.6\pm0.5$ & 0.022 & 0.278 & 0.1& 2.7& 3.3&4 \\
        2011dh & $55712.5\pm0.3$ & 0.001638 & $7.8\pm1.0$ & 0.035 & 0.05 & 0.07& 2.1&3.1 &5,6 \\
        2011fu & $55824.5\pm2.0$ & 0.01845 & $74.5\pm5.2$ & 0.068 & 0.035 & 0.15& 3.5& $\sim5.0$&7,8 \\
        2013df & $56447.8\pm0.9$ & 0.00239 & $21.4\pm3.0$ & 0.017 & 0.081 & 0.11& 1.5& $\lesssim3.0$&9,10 \\ 
        2015as & $57332.0\pm2.0$  & 0.0036 & $19.2\pm1.2$ & 0.008 & - &  0.08& 1.2& $\sim 2.7$& 11\\
        2016gkg & $57651.7^{\ddagger}$ & 0.0049 & $26.4\pm5.3$ & 0.0166 & 0.09 &  0.06& 1.6& 3.0& 12,13,14 \\
        2020acat & $59192.0\pm0.1$ & 0.007932 & $35.3\pm4.4$ & 0.0207 & - &  0.13& 1.7$-$2.3& $\sim4.5$& 15,16 \\
        2021bxu & $59246.3\pm0.4$ & 0.0178 & $72.0\pm5.0$ & 0.014 & - & 0.029 & 0.61& 2.1& 17 \\
        2022ngb & $59749.9\pm0.5$ & 0.00965 & $32.2\pm2.8$ & 0.085 & 0.085 & 0.045& 3.1& 4.7& This work \\
        2022crv & $59627.3\pm0.5$ &  0.008091  &  $34.4\pm2.4$ & 0.066 & 0.151  & 0.11& 3.9& 5.0& 18 \\
        2024abfo & $60628.3\pm0.1$ & 0.003512 & $10.9\pm1.3$ &0.0097 &   -    & 0.042 & 2.55$-$6.7& 4.1$-$8.2& 19,20 \\
        2024aecx &  $60660.0\pm0.8$ & 0.002665& $11.4\pm0.6$&  0.45 & - &  0.15& 0.70& 2.2& 21 \\
        \hline
    \end{tabular}
    }
    \label{tab_appSNIIbInfo}
    \vspace{2pt}
    \scriptsize 
    \raggedright
    
    \textbf{NOTE:} Sources: 1 = \citet{Richmond1994aj}; 2 = \citet{Barbon1995aaps}; 3 = \citet{Richmond1996aj}; 4 = \citet{Pastorello2008mnras}; 5 = \citet{Sahu2013mnras}; 6 = \citet{Ergon2015aap}; 7 = \citet{Kumar2013mnras}; 8 = \citet{Morales-Garoffolo2015mnras}; 9 = \citet{Morales-Garoffolo2014mnras}; 10 = \citet{VanDyk2014aj}; 11 = \citet{Gangopadhyay2018mnras}; 12 = \citet{Tartaglia2017apjl}; 13 = \citet{Arcavi2017apjl}; 14 = \citet{Bersten2018nat}; 15 = \citet{Medler2022MNRAS}; 16 = \citet{Ergon2024A&A}; 17 = \citet{Desai2023mnras}; 18 = \citet{Gangopadhyay2023ApJ}; 19 = \citet{deWet2025A&A...704A..89D}; 20 = \citet{Reguitti2025AA} ; 21 = \citet{zou2025arXiv}. $\ddagger$: The discovery epoch is very close to its explosion epoch that the uncertainty is negligible. *: Data are retrieved mostly from \citet{Barmentloo2024MNRAS}, considering having a better estimation and more precise.
\end{table*}
\begin{table*}[htbp]
    \centering
    \caption{Apparent light curve parameters of SN 2022ngb.}
    \begin{tabularx}{\textwidth}{ccYYYYY}
        \hline
        Filter & System & $t_{\mathrm{peak}}$ & $t_{\mathrm{rise}}$ & $m_{\mathrm{peak}}$ & $\gamma_{0-15}$ & $\gamma_{15-100}$ \\
        && [MJD] & [days] & [mag] & [mag (100 days)$^{-1}$] & [mag (100 days)$^{-1}$]\\
        \hline
        $B$&Vega& $ 59772.96 \pm 0.07 $& $ 23.06 \pm 0.50 $ & $ 17.66 \pm 0.01 $& $10.01\pm0.40$   & $0.77\pm0.06$ \\
        $g$&ab&   $ 59773.92 \pm 0.04 $& $ 24.01 \pm 0.50 $ & $ 17.22 \pm 0.01 $& $8.06\pm0.08$    & $0.98\pm0.04$ \\
        $c$$^{\dagger}$&ab&   -&                     - &                  -&                  $5.97\pm0.30$    & $1.43\pm0.11$ \\
        $V$&Vega& $ 59774.22 \pm 0.06 $& $ 24.32 \pm 0.50 $ & $ 16.74 \pm 0.01 $& $6.82\pm0.24$    & $1.25\pm0.03$ \\
        $r$&ab&   $ 59778.17 \pm 0.02 $& $ 28.27 \pm 0.50 $ & $ 16.53 \pm 0.01 $& $4.12\pm0.18$    & $1.55\pm0.02$ \\
        $o$&ab&   $ 59778.63 \pm 0.05 $& $ 28.73 \pm 0.50 $ & $ 16.44 \pm 0.01 $& $3.67\pm0.14$    & $1.64\pm0.05$ \\
        $i$&ab&   $ 59779.88 \pm 0.03 $& $ 29.97 \pm 0.50 $ & $ 16.35 \pm 0.01 $& $3.23\pm0.16$    & $1.89\pm0.03$ \\
        $z$&ab&   $ 59780.18 \pm 0.05 $& $ 30.27 \pm 0.50 $ & $ 16.33 \pm 0.01 $& -    & $1.94\pm0.03$ \\
        $J$$^{*}$&Vega& -&                     -&                   -&                  -    & $1.95\pm0.11$ \\
        $H$$^{*}$&Vega& -&                     -&                   -&                  -    & $1.58\pm0.21$ \\
        K$^{*}$&Vega& -&                     -&                   -&                  -    & $1.83\pm0.18$ \\
        \hline
    \end{tabularx}
    \label{tab_appapplc}
    \vspace{2pt}
    \scriptsize 
    \raggedright

    \textbf{NOTE:} 
    $\dagger$: The ATLAS \textit{c}-band data do not cover the maximum of SN~2022ngb, and therefore the fitting could not be performed. The corresponding decline rate is thus not reliable, as the late-phase data are of limited quality.  
    *: The NIR (\textit{JHK}) bands lack early phase and pre-maximum coverage, and the post-maximum data are also incomplete. Consequently, the fitting results should be treated with caution and considered less reliable.

\end{table*}
\begin{table*}[htbp]
\centering
\caption{Bolometric light curve fitting results of SN 2022ngb and other typical Type IIb SNe. }
\label{apptab_snlcfittingparams}
\footnotesize 
\resizebox{\textwidth}{!}{ 
\begin{tabular}{@{}l ccc ccc ccc ccc ccc@{}}
\toprule
& \multicolumn{3}{c}{$R_0$ ($10^{12}$ cm)} & \multicolumn{3}{c}{$M_{\mathrm{ej}}$ ($M_\odot$)} & \multicolumn{3}{c}{$M_{\mathrm{Ni}}$ ($M_\odot$)} & \multicolumn{3}{c}{$E_{\mathrm{tot}}$ ($10^{51}\,\mathrm{erg}$)} & \multicolumn{3}{c}{$E_\mathrm{k}/E_\mathrm{t}$} \\
\cmidrule(lr){2-4} \cmidrule(lr){5-7} \cmidrule(lr){8-10} \cmidrule(lr){11-13} \cmidrule(lr){14-16}
SN & core & shell & Arnett & core & shell & Arnett & core & shell & Arnett & core & shell & Arnett$^*$ & core & shell & Arnett \\
\midrule
1993J$^1$   & 0.35 & 30   & -                 & 2.15 & 0.1   & 1.9     & 0.1   & -    & $0.10 \pm 0.03$   & 3.7   & 0.8   & $1.3 \pm 0.3$    & 1.85  & 7    & -      \\
2008ax$^3$  & -    & -    & -                 & -    & -     & $2.7 \pm 0.5$     & -     & -    & $0.10 \pm 0.02$   &-  & -     & $1.2 \pm 0.5$                & -     & -    & -      \\
2011fu$^2$  & 0.35 & 13   & -                 & 2.2  & 0.12  & $3.5$             & 0.23  & -    & $0.15$            & 3.4   & 0.8   & $1.3$            & 2.4   & 1.67 & -      \\
2015as$^4$  & 0.2  & 0.5  & -                 & 1.2  & 0.1   & $2.2$             & 0.08  & -    & 0.07  & 1.14  & $0.58$            & 0.656 & 2.16 & $0.93$ & -\\
2020acat$^5$& -    & -    & -                 & -    & -     & $2.3 \pm 0.4$     & -     & -    & $0.13 \pm 0.03$   & -     & -     & $1.2 \pm 0.3$    & -     & -    & -      \\
2022crv$^6$&0.02&0.04$-$0.29&-&3.9&0.015$-$0.05&-&0.112&-&-&4.1&0.35&-&4.86&6.00&-\\
2024abfo$^7$&$\lesssim 0.1$&8.7$^{\dagger}$& - &2.55&$\lesssim 0.0065$& 4.1$-$6.7& 0.055& - & 0.042& 1.95$^{+}$ & - &1.5$-$6.7 & -& - & - \\
2024aecx$^{8}$&-&11.76$-$13.91&-&0.70&0.03$-$0.24&-&0.15&-&-&0.16$^{+}$&-&-&-&-&-\\
2022ngb & 0.10 & 0.3  & -   & 3.1 & 0.035 & 3.2     & 0.045 & -    & 0.045 & 1.94 & 0.47  & 1.34  & 2.11  & 5.8  & - \\
\bottomrule
\end{tabular}
}
\vspace{2pt}
\scriptsize 
\raggedright

\textbf{NOTE:} *: For all the Type IIb SNe listed in the table, the total energy of Arnett-like model here represent the kinetic energy. $\dagger$: The radius is resulting from the direct observation of the progenitor. +: for SN 2024abfo and SN 2024aecx, the initial thermal energy is not reported in two-components modeling, thus here is the kinetic energy instead. 1,2 = \cite{Nagy2016aap}, \cite{Medler2022MNRAS}, \cite{Kumar2013mnras}. 3 = \citet{Pastorello2008mnras}, \cite{Medler2022MNRAS}. 4 = \cite{Gangopadhyay2018mnras}. 5 = \cite{Medler2022MNRAS}. 6 = \cite{Gangopadhyay2023ApJ}. 7 = \cite{Reguitti2025AA}, \cite{deWet2025A&A...704A..89D}. 8 = \cite{zou2025arXiv}.
\end{table*}

\newpage
\section{Data description and reduction}
\label{app_datareduction}
\vspace{-0.5em}
\subsection{Photometric data reduction}
\vspace{-0.5em}
\label{app_subphotodatared}
We carried out comprehensive multiband follow-up observations for SN~2022ngb, in the Johnson-Cousins \textit{BV}, Sloan \textit{griz}, and NIR \textit{JHK} bands, soon after the SN~discovery. Optical and Near-infrared (NIR) data were obtained from the following facilities: 
The Las Cumbres Observatory (LCO) global telescopes located at Teide Observatory (TFN; specific the 1.0\,m telescope), Tenerife, Spain, with the fa11 and fa20 Sinistro cameras;
The 0.67\,m/0.92\,m Schmidt telescope with a Moravian camera at Padova Astronomical Observatory, Istituto Nazionale di Astrofisica (INAF), Asiago, Italy;
The 1.82\,m Copernico Telescope with the Asiago Faint Object Spectrograph and Camera (AFOSC), hosted by INAF -- Padova Astronomical Observatory, Asiago, Italy; 
The 2.0\,m Liverpool telescope (LT) equipped with the IO:O camera, located at Observatorio Roque de Los Muchachos, La Palma, Spain;
The 2.56\,m Nordic Optical Telescope (NOT), at Observatorio Roque de Los Muchachos, La Palma, Spain, with the Alhambra Faint Object Spectrograph and Camera (ALFOSC) and the Nordic Optical Telescope near-infrared Camera (NOTCam);
The 10.4\,m Gran Telescopio Canarias (GTC), at Observatorio Roque de Los Muchachos, La Palma, Spain, with the Optical System for Imaging and low-Intermediate-Resolution Integrated Spectroscopy (OSIRIS) instrument.

The initial reduction of all raw photometric data was performed using standard procedures within the \texttt{IRAF} \citep{Tody1986SPIE} environment. This process consisted of bias, overscan, trimming, and flat-fielding corrections. When multiple images in a band were observed in one night, we median-combined them into a single frame to increase the S/N. Subsequently, we performed the photometric measurements on these science images using the dedicated pipeline \texttt{ecsnoopy}\footnote{\texttt{ecsnoopy} is a package for SN~photometry using PSF fitting and/or template subtraction developed by E. Cappellaro. A package description can be found at \url{https://sngroup.oapd.inaf.it/ecsnoopy.html}}, which integrates several photometric packages, such as \texttt{SEXTRACTOR} \citep{Bertin1996A&AS} for source extraction, \texttt{DAOPHOT} \citep{Stetson1987PASP} for magnitude measurements, and  \texttt{HOTPANTS} \citep{Becker2015ascl} for image subtraction. Given that SN~2022ngb was contaminated by its host galaxy, we used the template-subtraction technique to remove the background contamination. Specifically, the SN instrumental magnitudes were measured using the Point Spread Function (PSF) fitting technique. The PSF model was constructed by fitting the profiles of isolated and nonsaturated stars in the SN field. The resulting modeled PSF profile was subtracted from the original frames, and the residuals at the position of SN were used to evaluate the fitting quality. If the SN is not detected, a magnitude limit was provided.

Then, the SN instrumental magnitudes were calibrated through the instrumental zero points (ZPs) and color terms (CTs), which were inferred from observations of standard stars on photometric nights.
Specifically, Johnsonn-Cousins magnitudes were determined from the \citet{Landolt1992AJ}
catalogue, while Sloan-filter magnitudes were calibrated using reference stars of the Pan-STARRS catalogue. In order to correct the ZPs on nonphotometric nights and improve the SN photometric calibration accuracy, we also applied the corrections of a local sequence of standard stars in the vicinity of the SN.

Photometric errors were estimated through artificial star tests. Several fake stars with known magnitudes were evenly placed near the position of SN~2022ngb. The  magnitudes of the simulated stars in the frame were subsequently measured with the PSF fits. The standard deviation of individual artificial star experiments was in quadrature combined with the PSF-fit and ZP-calibration errors, which finally provides the total photometric errors.

NIR data were reduced following similar procedures as the optical ones. Raw images were first reduced with bias correction, flat fielding, and distortion correction. We then median-combined several dithered science images to construct sky frames, which were combined to increase the S/N. In particular, the pre-reduction of NOT/NOTCam raw images was performed using a dedicated pipeline of NOTCam (version 2.5). Instrumental magnitudes were measured
through the PSF fits, and the final apparent magnitudes were calibrated using the Two Micron All Sky Survey \citep[2MASS;][]{Skrutskie2006AJ} catalogue. 

In addition, we also collect archival data from some public surveys, including ATLAS, ZTF, and Pan-STARRS. The ATLAS \textit{c-} and \textit{o}-band light curves are retrieved from the ATLAS forced photometry service \citep{Shingles2021TNSAN}. Using the script provided by \cite{Young2024zndo}, we stack the ATLAS photometric data in 1-day intervals for the early data and 3- or 5-day intervals for the late time data. The stacking uses the rolling-window technique to identify outliers and bin the data. ZTF g- and r-band data were obtained through ALeRCE \citep{Forster2021AJ} and Lasair \citep{Smith2019RNAAS} brokers. Pan-STARRS 2 (PS2; \citealp{Flewelling2018AAS, chambers2019arXiv}) images were processed with the Image Processing Pipeline (IPP; \citealt{Magnier2006amos, Magnier2020aas}) and calibrated to the Pan-STARRS DR1 catalogue \citep{Flewelling2020ApJS}. Additionally, some early data were retrieved from TNS AstroNotes\footnote{AstroNote 2022-131: \url{https://www.wis-tns.org/astronotes/astronote/2022-131}} in \textit{griz} bands \citep{Chen2022TNSAN}. 

Finally, the resulting optical and NIR magnitudes are reported in electronic form at the CDS. 
\vspace{-1.5em}
\subsection{Spectroscopic data reduction}
\vspace{-0.5em}
The spectroscopic data were obtained using the following telescopes and instruments: 
the 1.82\,m Copernico Telescope with AFOSC; the 2.56\,m NOT with ALFOSC; the 10.4\,m GTC with OSIRIS; the 3.6\,m Telescopio Nazionale \textit{Galileo} (TNG) at Observatorio Roque de Los Muchachos, La Palma, Spain, with its Low Resolution Spectrograph (LRS); the 11\,m Hobby-Eberly Telescope (HET; \citealt{Ramsey1998SPIE, Hill2021AJ}) at the McDonald Observatory, Texas, USA, with its Low Resolution Spectrograph 2 (LRS2; \citealt{Chonis2016SPIE}).

The raw spectroscopic data of SN~2022ngb were processed using standard routines within the \texttt{IRAF} package. Initial calibration steps, including bias subtraction, overscan correction, flat-fielding, and image trimming, were performed in an approach similar to the photometric data reduction described in the above Sect. \ref{app_subphotodatared}. The one-dimensional spectra were then optimally extracted from the processed two-dimensional frames. Wavelength calibration was achieved using arc-lamp observations taken during the same night, and flux calibration was performed using spectrophotometric standard stars. To verify the accuracy of the flux calibration, the calibrated spectra were checked and scaled against contemporaneous photometry of SN~2022ngb. Finally, the spectra of standard stars were also used to remove the telluric absorptions from the SN spectra.
A detailed description of the spectroscopic observations is reported in Table \ref{tab_2022ngbSpecInfo}.

\onecolumn

\section{Acknowledgements}
\begin{small}
We gratefully thank the anonymous referee for his/her insightful comments and suggestions that improved the paper.
We thank Luc Dessart for kindly providing the light curve and spectral models for SNe IIb, as well as for his valuable guidance and support in this work.
We thank L.-H. Li for helpful discussions and T.~Nagao for some data observations.
This work is supported by the National Key Research and Development Program of China (Grant No. 2024YFA1611603), the National Natural Science Foundation of China (NSFC, Grant Nos. 12303054, 12473047), the Yunnan Fundamental Research Projects (Grant Nos. 202401AU070063, 202501AS070078), the Yunnan Key Laboratory of Survey Science (No. 202449CE340002), and the International Centre of Supernovae, Yunnan Key Laboratory (No. 202302AN360001).
AP, AR, SB, EC, NER, LT, GV and IS acknowledge support from the PRIN-INAF 2022, "Shedding light on the nature of gap transients: from the observations to the models". AR also acknowledges financial support from the GRAWITA Large Program Grant (PI P. D'Avanzo).
N.E.R. also acknowledges support from the Spanish Ministerio de Ciencia e Innovaci\'on (MCIN) and the Agencia Estatal de Investigaci\'on (AEI) 10.13039/501100011033 under the program Unidad de Excelencia Mar\'ia de Maeztu CEX2020-001058-M.
A. F. acknowledges funding by the European Union - NextGenerationEU RFF M4C2 1.1 PRIN 2022 project "2022RJLWHN URKA" and
by INAF 2023 Theory Grant ObFu 1.05.23.06.06 "Understanding R-process \& Kilonovae Aspects (URKA)".
EC acknowledges support from MIUR, PRIN 2020 (METE, grant 2020KB33TP).
BK is supported by the ``Special Project for High-End Foreign Experts", Xingdian Funding from Yunnan Province.
S. Mattila acknowledges financial support from the Research Council of Finland project 350458.
A.F. acknowledges the support by the State of Hesse within the Research Cluster ELEMENTS (Project ID 500/10.006).
S. Moran is funded by Leverhulme Trust grant RPG-2023-240.
M.D. Stritzinger is funded by the Independent Research Fund Denmark (IRFD, grant number 10.46540/2032-00022B).
T.K. acknowledges support from the Research Council of Finland project 360274.
JV is supported by NKFIH-OTKA Grant K142534.
S.-P. Pei is supported by the  Science and Technology Foundation of Guizhou Province (QKHJC-ZK[2023]442). 
T.M.R is part of the Cosmic Dawn Center (DAWN), which is funded by the Danish National Research Foundation under grant DNRF140. T.M.R acknowledges support from the Research Council of Finland project 350458.
M.D.S. is funded by the Independent Research Fund Denmark (IRFD, grant number  10.46540/2032-00022B) and by an Aarhus University Research Foundation Nova project (AUFF-E-2023-9-28).
Y.-J. Yang is supported by the National Natural Science Foundation of China (Grants No. 12305066).
S.M. acknowledges financial support from the Research Council of Finland project 350458.

We acknowledge the support of the staffs of the various observatories at which data were obtained.
Based on observations made with the Nordic Optical Telescope (NOT), owned in collaboration by the University of Turku and Aarhus University, and operated jointly by Aarhus University, the University of Turku, and the University of Oslo, representing Denmark, Finland, and Norway, the University of Iceland, and Stockholm University at the Observatorio del Roque de los Muchachos, La Palma, Spain, of the Instituto de Astrofisica de Canarias. Observations from the NOT were obtained through the NUTS2 collaboration which is supported in part by the Instrument Centre for Danish Astrophysics (IDA), and the Finnish Centre for Astronomy with ESO (FINCA) via Academy of Finland grant nr 306531. The data presented here were obtained in part with ALFOSC, which is provided by the Instituto de Astrofisica de Andalucia (IAA) under a joint agreement with the University of Copenhagen and NOTSA.
The Liverpool Telescope is operated on the island of La Palma by Liverpool John Moores University in the Spanish Observatorio del Roque de los Muchachos of the Instituto de Astrofisica de Canarias with financial support from the UK Science and Technology Facilities Council.
The Italian Telescopio Nazionale \textit{Galileo} (TNG) is operated on the island of La Palma by the Fundaci\'on Galileo Galilei of the INAF (Istituto Nazionale di Astrofisica), at the Spanish Observatorio del Roque de los Muchachos of the Instituto de Astrof\'isica de Canarias.
Based on observations collected at Copernico and Schmidt telescopes (Asiago, Italy) of the INAF -- Osservatorio Astronomico di Padova.
Based on observations made with the Gran Telescopio Canarias (GTC), (Programs GTCMULTIPLE2A-22A and GTCMULTIPLE2G-22B; PI: Nancy Elias-Rosa) installed at the Spanish Observatorio del Roque de los Muchachos of the Instituto de Astrof\'{i}sica de Canarias, on the island of La Palma.
This work makes use of data from the Las Cumbres Observatory (LCO) Network and the Global Supernova Project. The LCO team is supported by U.S. NSF grants AST-1911225 and AST-1911151, and NASA.
Based on observations obtained with the Hobby-Eberly Telescope (HET), which is a joint project of the University of Texas at Austin, the Pennsylvania State University, Ludwig-Maximillians-Universitaet Muenchen, and Georg-August Universitaet Goettingen. The HET is named in honor of its principal benefactors, William P. Hobby and Robert E. Eberly. The Low Resolution Spectrograph 2 (LRS2) was developed and funded by the University of Texas at Austin McDonald Observatory and Department of Astronomy, and by Pennsylvania State University. We thank the Leibniz-Institut fur Astrophysik Potsdam (AIP) and the Institut fur Astrophysik Goettingen (IAG) for their contributions to the construction of the integral field units. We acknowledge the Texas Advanced Computing Center (TACC) at The University of Texas at Austin for providing high performance computing, visualization, and storage resources that have contributed to the results reported within this paper.

This work has made use of data from the Asteroid Terrestrial-impact Last Alert System (ATLAS) project. The Asteroid Terrestrial-impact Last Alert System (ATLAS) project is primarily funded to search for near earth asteroids through NASA grants NN12AR55G, 80NSSC18K0284, and 80NSSC18K1575; byproducts of the NEO search include images and catalogs from the survey area. This work was partially funded by Kepler/K2 grant J1944/80NSSC19K0112 and HST GO-15889, and STFC grants ST/T000198/1 and ST/S006109/1. The ATLAS science products have been made possible through the contributions of the University of Hawaii Institute for Astronomy, the Queen's University Belfast, the Space Telescope Science Institute, the South African Astronomical Observatory, and The Millennium Institute of Astrophysics (MAS), Chile. 
The Pan-STARRS2 Surveys (PS2) and the PS2 public science archive have been made possible through contributions by the Institute for Astronomy, the University of Hawaii, the Pan-STARRS Project Office, the Max-Planck Society and its participating institutes, the Max Planck Institute for Astronomy, Heidelberg and the Max Planck Institute for Extraterrestrial Physics, Garching, The Johns Hopkins University, Durham University, the University of Edinburgh, the Queen's University Belfast, the Harvard-Smithsonian Center for Astrophysics, the Las Cumbres Observatory Global Telescope Network Incorporated, the National Central University of Taiwan, the Space Telescope Science Institute, the National Aeronautics and Space Administration under Grant No. NNX08AR22G issued through the Planetary Science Division of the NASA Science Mission Directorate, the National Science Foundation Grant No. AST-1238877, the University of Maryland, Eotvos Lorand University (ELTE), the Los Alamos National Laboratory, and the Gordon and Betty Moore Foundation.
The Zwicky Transient Facility (ZTF) is supported by the National Science Foundation under Grants No. AST-1440341 and AST-2034437 and involves a collaboration that includes current partners such as Caltech, IPAC, the Oskar Klein Center at Stockholm University, the University of Maryland, the University of California, Berkeley, the University of Wisconsin-Milwaukee, the University of Warwick, Ruhr University, Cornell University, Northwestern University, and Drexel University. Operations are conducted by COO, IPAC, and UW.

\end{small}
\onecolumn

\twocolumn

\end{appendix}

\end{document}